%% file: main.tex
\documentclass[letterpaper,twocolumn,10pt]{article}
\usepackage{usenix-2020-09}

\usepackage[]{hyperref}
\usepackage{xspace}
\usepackage{graphicx}
\usepackage{subcaption}
\usepackage{lipsum}
\usepackage{enumerate}
\usepackage{amsmath}
\usepackage{amsthm}
\usepackage{amsfonts}
\usepackage{fontawesome}
\usepackage[dvipsnames]{xcolor}
\usepackage{tikz}
\usetikzlibrary{fit, backgrounds, positioning, shapes.geometric, arrows.meta}
\usepackage{pgfplots}
\pgfplotsset{compat=1.18}
\usepackage{pifont}
\usetikzlibrary{patterns}
\usepackage{tabularx}
\usepackage{booktabs}
\usepackage{algorithm}
\usepackage{algpseudocode}
\usepackage{titling}

\newcommand{\sys}{Perseus\xspace}
\providecommand{\ie}{\emph{i.e.,} }
\providecommand{\eg}{\emph{e.g.,} }
\providecommand{\vs}{vs. }
\providecommand{\myparab}[1]{\noindent\textbf{#1}}

\providecommand{\myparab}[1]{\vspace{1pt}\noindent\textbf{#1} }
\providecommand{\alltoall}{\textsc{AllToAll}\xspace}

\providecommand{\PUT}{\textsc{Put}\xspace}
\providecommand{\puts}{\textsc{PUT}s\xspace}
\providecommand{\fence}{\textsc{Fence}\xspace}
\providecommand{\fences}{\textsc{Fence}s\xspace}
\providecommand{\signal}{\textsc{Signal}\xspace}
\providecommand{\signals}{\textsc{Signal}s\xspace}
\providecommand{\putsignal}{\textsc{Put-with-Signal}\xspace}

\newenvironment{packeditemize}{\begin{list}{$\bullet$}{\setlength{\itemsep}{0.1pt}\addtolength{\labelwidth}{0pt}\setlength{\leftmargin}{\labelwidth}\setlength{\listparindent}{\parindent}\setlength{\parsep}{1pt}\setlength{\topsep}{0pt}}}{\end{list}}

\usepackage[small,compact]{titlesec}
\date{}

\begin{document}
\setlength{\droptitle}{-2em}   
\posttitle{\par\end{center}}

\title{\Large \bf Eliminating Hidden Serialization in Multi-Node Megakernel Communication}

\author{
{\rm Byungsoo Oh}\\
Cornell University
\and
{\rm Rachee Singh}\\
Cornell University
}

\maketitle

\input{tex/abstract.tex}
\input{tex/intro.tex}
\input{tex/background.tex}
\input{tex/motivation.tex}
\input{tex/design.tex}
\input{tex/implementation.tex}
\input{tex/eval.tex}
\input{tex/related.tex}
\input{tex/discussion.tex}
\input{tex/conclusion.tex}

\bibliographystyle{plain}
\bibliography{reference.bib}

\appendix
\input{tex/appendix.tex}

\end{document}

%% file: tex/abstract.tex
\begin{abstract}
    Recent megakernel designs for Mixture-of-Experts (MoE) inference fuse expert computation with fine-grained, GPU-initiated communication into a single persistent GPU kernel, and outperform collective-based MoE on a single node by overlapping data transfer with compute at tile granularity. This benefit does not carry over cleanly to multi-node inference, where experts span many nodes connected by an RDMA fabric. Communication-bound MoE models regress by up to $10\times$ on 8 nodes, and the regression worsens with node count. We trace this regression to hidden serialization in proxy-based RDMA transports. The ordering requirement between each tile transfer and its completion signal forces a fence that drains the NIC pipeline, and its cost grows with the number of concurrent transfers. As a result, models whose per-expert compute is too small to absorb this inflated network latency expose communication on the critical path. We present \sys, which eliminates this serialization through two techniques. \emph{Decoupled signaling} batches fences at per-destination granularity, reducing fence count by $8\times$. \emph{NIC-side ordering} replaces proxy stalls with hardware fence flags, so the proxy never blocks. On proxy-based transports, \sys achieves up to 10.3$\times$ end-to-end speedup. \sys on IBRC matches or exceeds IBGDA GPU-direct by up to 1.2$\times$, which shows that serialization, rather than the choice between proxy-based and GPU-direct transport, is what bounds multi-node megakernel performance.
\end{abstract}

%% file: tex/intro.tex
\section{Introduction}
\label{sec:intro}

Mixture-of-Experts (MoE) has become the dominant architecture for large language models. Frontier models including Qwen3~\cite{qwen3}, DeepSeek-V3~\cite{deepseek-v3}, and Llama4~\cite{llama4} all adopt it. MoE activates only a subset of parameters per token, which lets models scale parameter counts without a proportional increase in inference compute. This property has enabled the recent progress in model quality under practical serving budgets. This architectural efficiency comes with an underlying challenge. Expert parallelism partitions the experts in an MoE model across GPUs, so every token must travel to the GPU that holds its assigned expert and its output must return before the next layer can proceed~\cite{gshard,deepspeed-moe}. The resulting all-to-all exchange lies on the inference critical path, and its cost governs how MoE systems scale in production.

Existing MoE implementations shuffle tokens with  all-to-all collective communication primitives from libraries like NCCL~\cite{nccl}, invoking one collective before expert computation and another after~\cite{gshard,deepspeed-moe}. A collective is a bulk synchronous operation that requires all GPUs to receive tokens before any one can begin computation. This layer-wide serialization between communication and computation rules out overlap at any finer granularity, which is where most of the recoverable performance lies~\cite{flashmoe,mpk,triton-dist,parallelkittens}.

A recent line of work replaces the collective all-to-all with asynchronous point-to-point transfers issued from inside the MoE GPU kernel~\cite{flashmoe,mpk,triton-dist,parallelkittens}. In this \emph{megakernel} design, the unit of work is a \emph{tile} of expert output. As soon as a GPU produces a tile, a thread on that GPU issues a one-sided asynchronous transfer to the destination GPU and a signal that marks the tile as ready. The receiver polls the signal, and any tile whose signal has arrived can enter the next stage of computation while the rest of the exchange is still in flight. GPU-initiated communication frameworks, including NVSHMEM~\cite{nvshmem} and the NCCL device APIs~\cite{nccl-gin,nccl-ep}, expose the primitives that make this pattern expressible inside a kernel. Within a single multi-GPU node, the megakernel design significantly outperforms collective-based MoE~\cite{flashmoe}.

\begin{figure}[t]
    \centering
    \includegraphics[width=\linewidth]{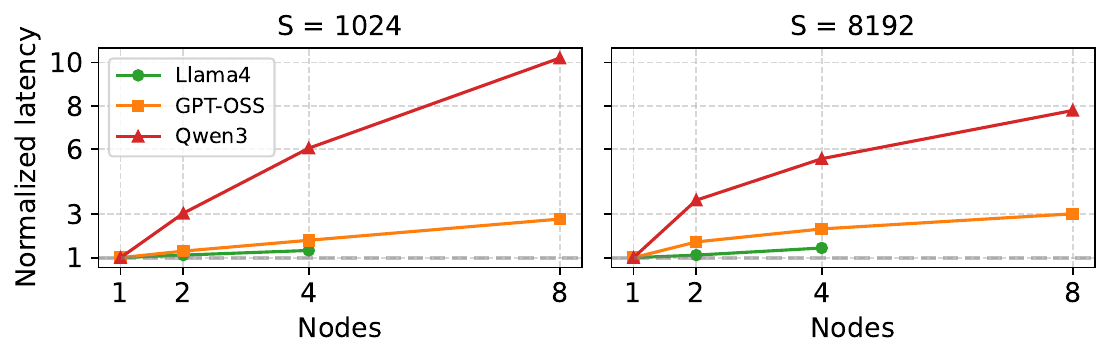}
    \vspace{2pt}
    \begin{minipage}{0.48\linewidth}
        \centering
        \includegraphics[width=\linewidth]{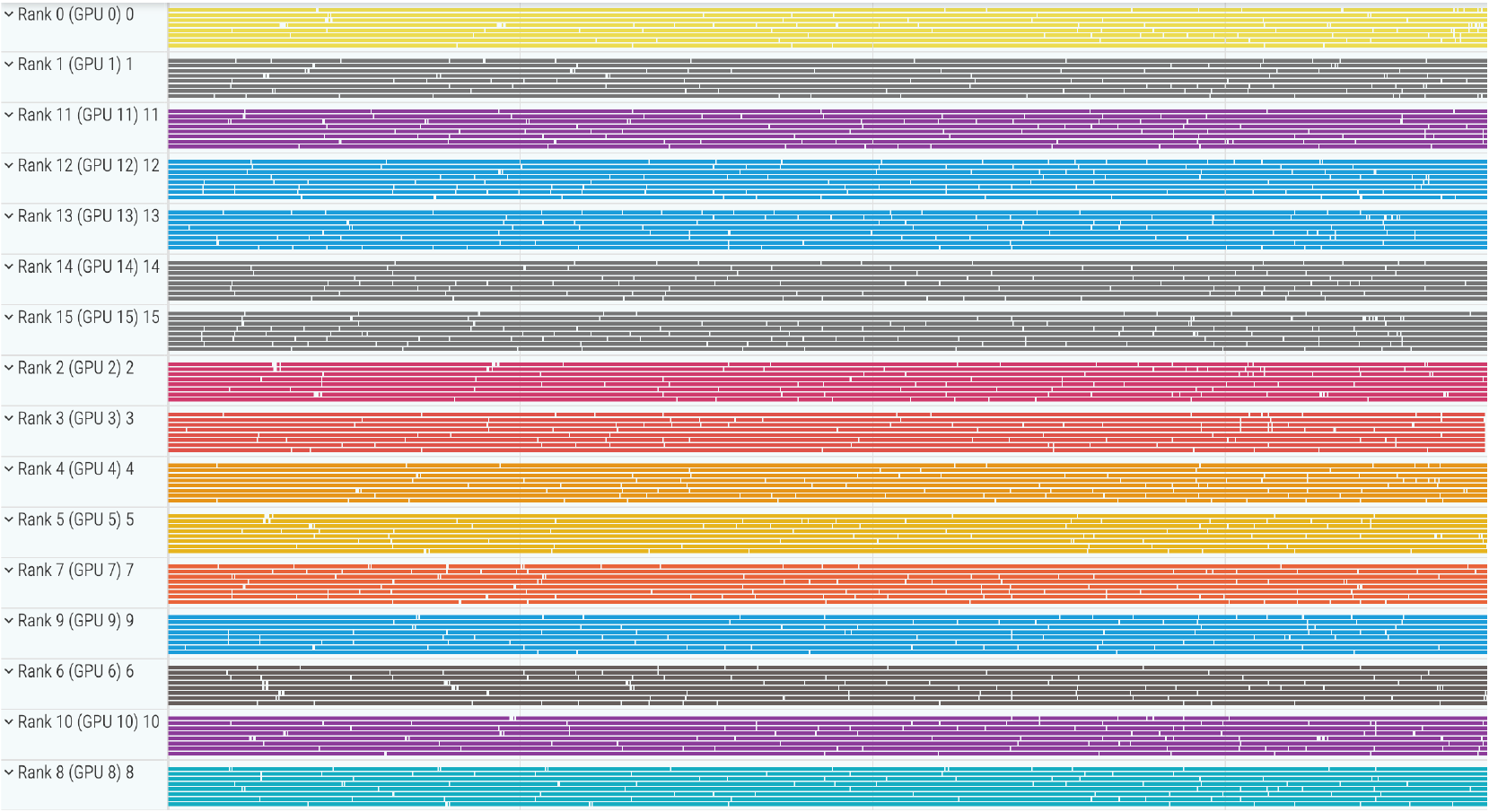}
        \small (a) Llama4
    \end{minipage}
    \hfill
    \begin{minipage}{0.48\linewidth}
        \centering
        \includegraphics[width=\linewidth]{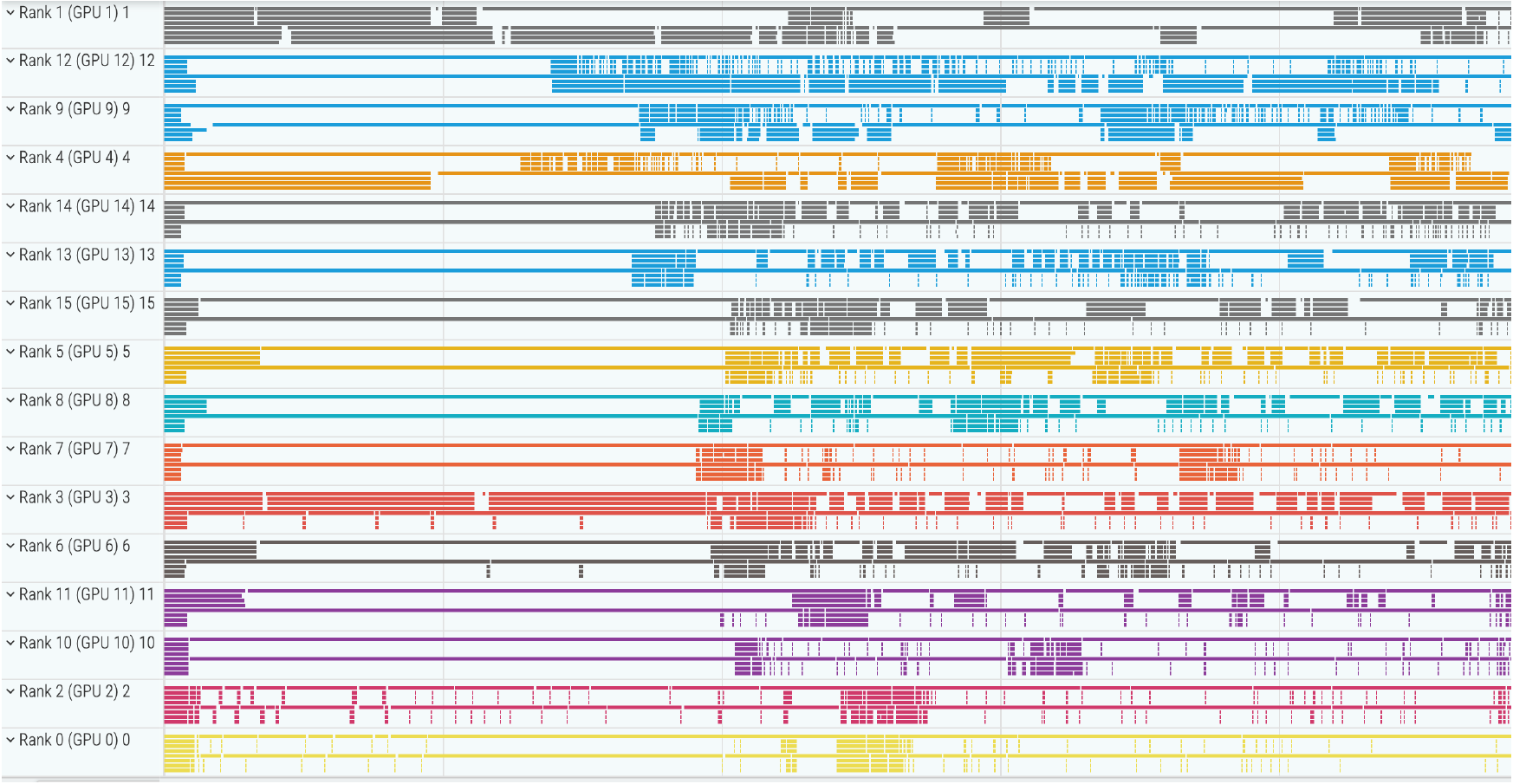}
        \small (b) Qwen3
    \end{minipage}
    \caption{Top: Weak scaling (ideal = 1.0$\times$). Llama4 uses 16 experts, limiting EP-only scaling to 4 nodes (16 GPUs). Bottom: GPU streaming multiprocessor (SM) traces at 4 nodes. Llama4 shows dense overlap; Qwen3 shows frequent stalls due to delayed signals.}
    \label{fig:motivation-weak-scaling}
    \vspace{-0.5mm}
\end{figure}

Frontier-scale MoE models require an expert-parallel degree larger than the GPU count of a single node, and these models therefore place experts across multiple nodes connected by a network fabric. Extending MoE megakernels to this setting is harder than prior work suggests. Figure~\ref{fig:motivation-weak-scaling} reports the end-to-end latency of forward pass on state-of-the-art MoE megakernels sharded across increasing node counts in our experiments. Performance degrades relative to single-node baselines and continues to worsen as node count grows. The degradation depends on model architecture. Compute-bound models like Llama4 scale near-linearly with node count, whereas communication-bound models like Qwen3-30B run up to 10$\times$ slower on 8 nodes than the single-node trend would predict.

The root cause of this regression is the inflated latency of communication over the inter-node RDMA transport. Megakernels hide communication behind computation at tile granularity, and this hiding works only when the per-tile compute runs long enough to absorb the per-tile network latency. When the transport itself becomes slow, the budget for this overlap shrinks. Figure~\ref{fig:motivation-weak-scaling}b makes this visible. Qwen3, whose expert compute is small, has little headroom to absorb inflated transport latency, and its streaming multiprocessor (SM) trace shows frequent stalls where processors wait for signals that arrive late. Llama4, with much larger per-expert compute, retains enough slack to hide the same transport cost, and its trace remains densely packed.

We find that the transport slowdown happens due to hidden serialization in the RDMA path. Recall that megakernels signal the remote GPU after every tile transfer so the receiver can begin computing immediately. Correct ordering requires a \emph{fence} between each transfer and its signal so the signal becomes visible only after the data. On proxy-based transports, where a CPU thread mediates GPU-initiated RDMA operations, this fence drains every in-flight data transfer from the NIC pipeline before the proxy submits the signal. A single dispatch phase in MoE models triggers one drain per expert transfer, which amounts to 96 drains in our Qwen3-30B configuration, and the per-drain cost grows with node count as concurrent transfers accumulate. We show that signaled transfer throughput falls to as little as 2\% of its unsignaled counterpart at 96 concurrent transfers across 8 nodes.

\begin{figure}[t]
\centering
\includegraphics[width=\linewidth]{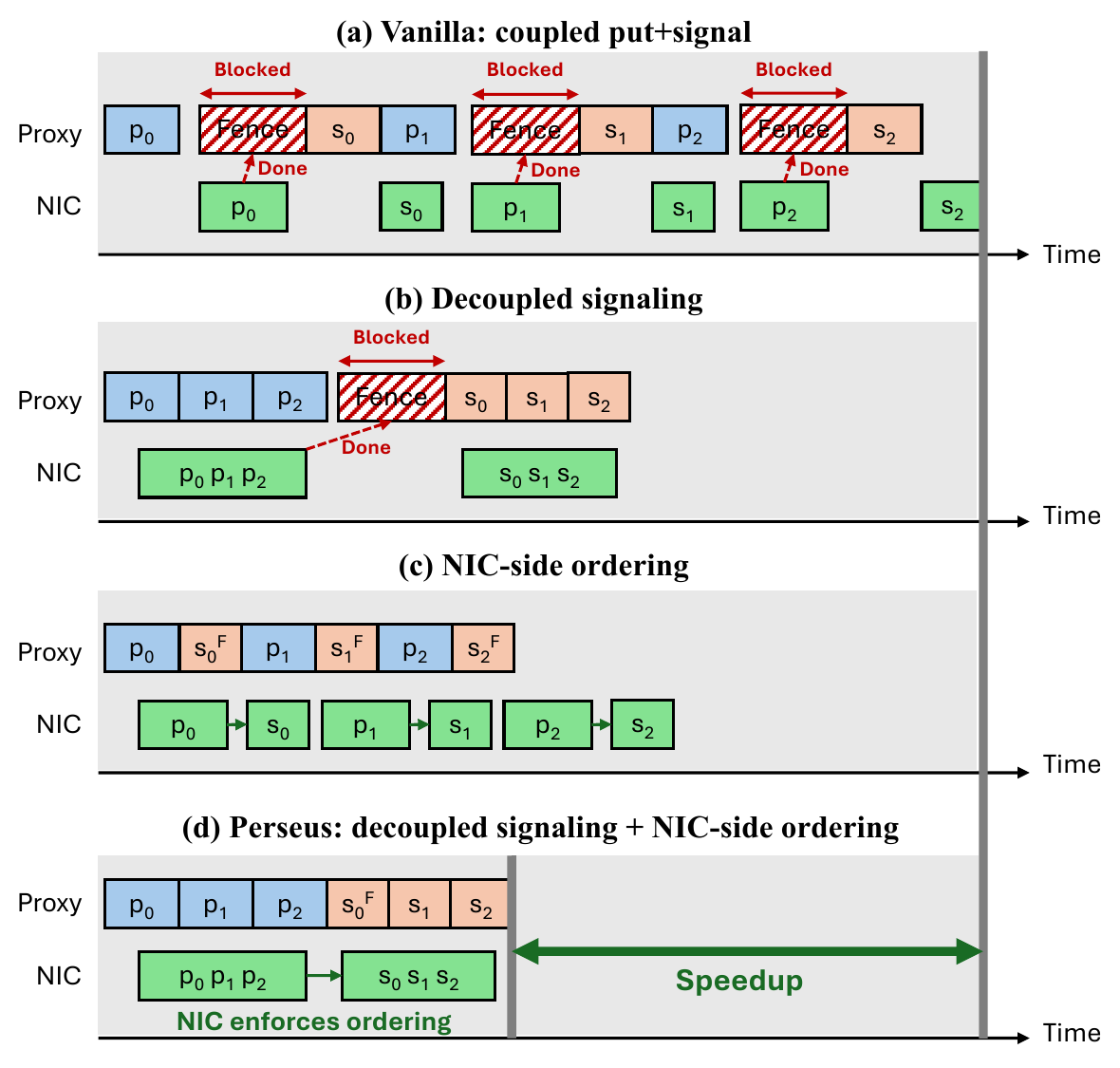}
\caption{\small{Proxy--NIC interaction for 3 signaled transfers
(p$_i$ = PUT, s$_i$ = signal).
(a)~\textbf{Vanilla}: each coupled put+signal inserts a proxy fence (3 proxy stops).
(b)~\textbf{Decoupled signaling}: PUTs submitted back-to-back; one proxy fence precedes all signals (1 proxy stop).
(c)~\textbf{NIC-side ordering}: proxy delegates ordering to NIC via per-signal fence flags (s$_i$\textsuperscript{F}); proxy never blocks but each fence flag serializes the NIC pipeline (0 proxy stops, 3 NIC stalls).
(d)~\textbf{\sys (both)}: only the first signal carries the fence flag (0 proxy stops, 1 NIC stall).}}
\label{fig:proxy-nic-timeline}
\end{figure}

We present \emph{\sys}\footnote{\sys is named after the Greek hero who slays Medusa, whose gaze freezes everything to stone.}, a system that eliminates hidden serialization in megakernel communication while preserving tile-level overlap of compute and communication. \sys introduces two complementary techniques, one at the megakernel layer and one at the transport layer (Figure~\ref{fig:proxy-nic-timeline}). \sys's \textbf{Decoupled signaling} separates data transfer from the signal that announces its arrival. GPU thread blocks issue transfers back-to-back without blocking, and a designated leader per group issues a single fence followed by all signals for that group. Grouping signals by destination GPU reduces the number of fences from one per expert to one per remote GPU---an 8$\times$ reduction in our running Qwen3 example---while preserving transfer concurrency. \textbf{NIC-side ordering} addresses the remaining per-fence cost at the transport level. Rather than blocking the proxy to drain the NIC pipeline, \sys attaches a hardware fence flag to the signal submission and delegates ordering enforcement to the NIC. The proxy never blocks, and only the first signal per group carries the flag.

We evaluate \sys on two platforms: Perlmutter (A100, Slingshot-11, up to 64 GPUs) and a commercial GPU cloud (H100, ConnectX-7, up to 32 GPUs). Our evaluation spans three RDMA transport backends (Libfabric proxy, IBRC proxy, IBGDA GPU-direct), three MoE models that span the compute-to-communication spectrum (Qwen3-30B, GPT-OSS-120B, and DeepSeek-v3), and sequence lengths from 256 to 64K tokens. On proxy-based transports, \sys achieves up to 10.3$\times$ end-to-end speedup on Libfabric and up to 2.47$\times$ on IBRC. \sys on IBRC matches or exceeds vanilla IBGDA GPU-direct by up to 1.2$\times$, which shows that serialization, rather than the choice between proxy-based and GPU-direct transport, is what bounds multi-node megakernel performance. \sys's transport-level optimization is portable to various ML frameworks. We apply it to Triton-Distributed's \alltoall benchmark without any application-level changes, where \sys achieves up to 79$\times$ speedup by eliminating 99\% of synchronization overhead.

%% file: tex/background.tex
\section{The Rise of Distributed Megakernels}
\label{sec:background}

Distributed Mixture-of-Experts (MoE) inference exposes the inefficiencies of the CPU-driven execution model that ML frameworks rely on. In this model, the host CPU launches GPU tasks, called kernels, one at a time onto GPU streams, and each kernel runs to completion across all of the GPU's streaming multiprocessors (SMs) before the next kernel begins. Inter-GPU communication uses host-initiated collectives that impose a global synchronization barrier across every participating GPU. These coarse-grained synchronization points in both compute and communication kernels leave SMs idle for large fractions of time. A new class of \emph{megakernel} designs has emerged in response. A megakernel fuses an entire MoE layer, or even an entire model, into one persistent GPU kernel and overlaps compute with GPU-initiated communication at tile granularity~\cite{flashmoe,mpk,triton-dist,comet,hazy-megakernel}. This section provides background on how megakernels address the performance bottlenecks of distributed MoE inference.

\subsection{Mixture-of-Experts and expert parallelism}
\label{sec:bg:moe}

A standard Transformer block in a large-language model consists of a self-attention module followed by a dense feed-forward network (FFN)~\cite{transformer}. A Mixture-of-Experts (MoE) block replaces the single FFN with a collection of $E$ identically shaped FFNs, called \emph{experts}, together with a lightweight gating network that selects a small subset of experts for each token~\cite{moe,gshard,deepseek-v3,llama4,qwen3}. For each token, the gate produces affinity scores over all experts and routes the token to its top-$k$ experts, with $k$ much smaller than $E$. The MoE layer's output is a weighted combination of the selected experts' outputs. This conditional computation decouples parameter count from per-token FLOPs, so adding experts grows model capacity without proportionally growing inference cost. Most modern frontier models use the MoE architecture \eg DeepSeek-V3 uses $E{=}256$ experts with top-8 routing~\cite{deepseek-v3}, Qwen3-235B uses $E{=}128$~\cite{qwen3}, and Llama4 Maverick uses $E{=}128$~\cite{llama4}.

\myparab{Expert parallelism (EP).}
The aggregate expert weights in frontier MoE models far exceed the memory of a single GPU, so experts are partitioned across GPUs using \emph{expert parallelism}~\cite{gshard,deepspeed-moe}. Let $P$ denote the number of GPUs (processing elements or PEs) in the expert-parallel group; each PE hosts $E/P$ experts. Since the gate assigns tokens dynamically at runtime, any PE may send tokens to any other PE. Each MoE layer therefore consists of four steps:
(1) a local \emph{gate} compute that assigns each token to $k$ experts, (2) a \emph{dispatch} \alltoall that moves each token to the PE(s) hosting its selected experts, (3) an \emph{expert compute} step, two General Matrix Multiplications (GEMMs) and an activation, on the received tokens, and (4) a \emph{combine} \alltoall that returns expert outputs to the originating PEs for weighted aggregation. Dispatch and combine are on the critical path of every layer, and their latency scales with both $E$ and $P$.

\subsection{The conventional CPU-driven execution model}
\label{sec:bg:kernel-model}

GPU workloads in modern ML frameworks like PyTorch~\cite{pytorch} are expressed as sequences of short-lived kernels submitted by the host CPU to GPU streams. PyTorch code for a matrix multiplication, an activation, a reduction, compiles into one or more GPU kernel launches. The host CPU enqueues each kernel into a GPU stream and returns immediately, while the GPU executes enqueued kernels in FIFO order. This execution model is attractive because it is composable (operators can be combined as library calls) and debuggable (each kernel is an independent failure unit). For dense models at modest scale, it is also efficient, because kernels are large enough that the overhead of launching kernels amortizes.

Two properties of the CPU-driven model limit MoE inference performance. The first is the launch overhead that the model imposes on every kernel. A single MoE layer, implemented naively, dispatches hundreds of short-lived kernels. Recent work reports up to 550 kernel launches per layer for DeepSpeedMoE and 261 for Megatron-LM~\cite{flashmoe}. Each launch carries a fixed submission cost from the host, and when kernels are this short and this numerous, the host may not keep the GPU's stream populated ahead of consumption causing GPU SMs to be idle. The second is the all-or-nothing dependency that the CPU-driven model enforces at collective boundaries. Collective primitives including \alltoall complete as a single unit, so the entire dispatch must finish before \emph{any} token can enter expert compute, and the entire expert compute must finish before combine can begin. 

These two properties compound for communication-bound workloads like MoE inference. The critical path is dominated by collectives, and the collectives are themselves punctuating hundreds of short kernels whose launch gaps lead to idle GPU SMs. Recent measurements on two A100 GPUs show GPUs idle 60--90\% of the time during MoE forward passes, with kernel launch gaps and non-overlapped \alltoall each contributing a large fraction of the idleness~\cite{flashmoe}.

\begin{figure}[t]
\centering
\includegraphics[width=0.9\linewidth]{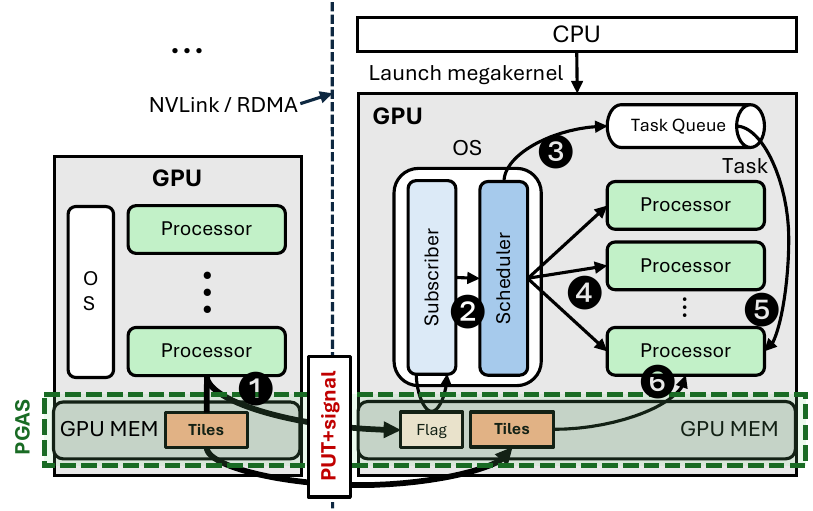}
\caption{Megakernel execution model. Each GPU runs a persistent kernel launched once by the host. CTAs specialize as processors or OS (subscriber + scheduler). Symmetric memory on both GPUs forms a PGAS address space (green dashed) enabling one-sided communication. Workflow: \ding{182} processor on one GPU issues \putsignal to peer's symmetric memory; \ding{183} subscriber polls flag and notifies scheduler; \ding{184}--\ding{185} scheduler enqueues and assigns tasks; \ding{186}--\ding{187} processor dequeues, reads the arrived tile, and computes. The left GPU's OS internals, task queue, and host CPU are omitted for clarity; both GPUs run identical megakernel code.}
\label{fig:bg-megakernel}
\end{figure}

\subsection{Megakernels: persistent GPU kernels}
\label{sec:bg:megakernel}

Megakernels attack both costs of the CPU-driven execution model simultaneously by fusing an entire subgraph of operators (each conventionally a separate kernel) into a single, persistent GPU kernel that never returns control to the host mid-computation~\cite{flashattention,flashmoe,mpk,hazy-megakernel,comet,triton-dist}. For instance, recent work replaces the 260--550 kernels of a conventional MoE layer with only a single kernel~\cite{flashmoe}.

\myparab{Tile-level parallelism.}
A megakernel operates at the granularity of \emph{tiles}, fixed-size sub-matrices of the input chosen at compile time (\eg $128{\times}64$~\cite{flashmoe}), rather than over whole input tensors. After the gate routes each token to its top-$k$ experts, megakernels group tokens into tiles, and each tile traverses the dependency graph consisting of a dispatch, a GEMM, an activation, a second GEMM, combine, and accumulation. Megakernels track dependencies per tile, so a tile advances to its next stage the moment its inputs are ready, regardless of the progress of any other tile. 

\myparab{Managing tile dependencies.}
Since a megakernel never returns control to the host, it must manage its own execution on the GPU. It does this by specializing cooperative thread arrays (CTAs), the parallelizable unit of GPU computation, into distinct roles (Figure~\ref{fig:bg-megakernel}). Most CTAs serve as \emph{processors} that execute computation (\eg GEMMs) on individual tiles. A small number of CTAs, or warps within a CTA, play the role of an \emph{operating system} that orchestrates the dependency pipeline. This operating system includes a \emph{subscriber} that decodes tile-arrival notifications from peer GPUs and a \emph{scheduler} that maintains a ready queue of tile-level tasks and assigns them to idle processors~\cite{flashmoe,mpk}. Because the scheduler runs on the GPU and observes readiness directly, a tile transitions from \emph{arrived} to \emph{under compute} without a host round-trip. The resulting megakernel designs report over 93\% SM utilization under this organization, compared to 9--60\% for CPU-driven baselines~\cite{flashmoe}.

\myparab{GPU-initiated communication.}
A megakernel runs entirely on the GPU and does not return control to the host until completion, so host-launched collectives including \alltoall are unreachable from inside the megakernel~\cite{gpu-centric}. Even if they were reachable, tile-level scheduling is only useful if communication can be issued and observed at tile granularity, yet a collective's unit is an entire layer of the model and its completion is a global barrier, which is the coarse-grained synchronization the megakernel was designed to eliminate. Megakernels therefore require communication that is both \emph{device-initiated} and \emph{one-sided}. Device-initiated means that a GPU thread, rather than the host CPU, issues the data transfer, which is what allows communication to be expressed from inside a persistent kernel. One-sided means that a sender can write directly into the receiver's memory without the receiver posting a matching receive, which is what allows a producer CTA in the megakernel to hand a tile to its consumer at the moment the tile is ready, rather than waiting for the consumer to reach a rendezvous point. 

\myparab{Global addressing enables one-sided communication.}
One-sided communication requires the sender to know where on the receiver GPU's memory the data should land, because the receiver is not involved in the transfer and cannot supply a destination. The Partitioned Global Address Space (PGAS) model~\cite{pgas} gives the sender this knowledge. Every PE allocates a \emph{symmetric} memory region at the same virtual offset, so the union of these regions forms a single logical address space in which a sender can name any location on any remote PE using ordinary pointer arithmetic, with no prior coordination. GPU-initiated communication libraries including NVSHMEM~\cite{nvshmem}, rocSHMEM~\cite{rocshmem}, NCCL device APIs~\cite{nccl-gin,nccl-ep}, NCCLX~\cite{ncclx}, and PyTorch Symmetric Memory~\cite{torch-symm-mem} expose this model to code, giving a CTA on a PE the ability to write into a symmetric buffer on another PE with a single library call.

\myparab{Signaling in one-sided communication.}
The communication primitive that binds PGAS to tile-level scheduling in megakernels is called \emph{put-with-signal}. A processor CTA that has finished staging tokens for one remote expert calls \(\texttt{putmem\_signal\_nbi}()\) (Figure~\ref{fig:bg-megakernel} \ding{182}), which performs two operations that appear atomic from the programmer's point of view: (1) a non-blocking write of the tokens into the remote symmetric buffer or \PUT, and (2) a write to a small flag variable on the remote PE after the data is guaranteed to be visible there or \signal. The remote subscriber CTA polls the flag; when the flag flips, the subscriber enqueues the dependent compute tasks and processor CTAs pick them up. This entire exchange takes place without host participation or collective barriers, and the sender CTA moves on to its next task the instant the call returns, making this approach highly efficient.

%% file: tex/motivation.tex
\section{Scaling Megakernels Beyond a Single Node}
\label{sec:bg:multinode}
State-of-the-art distributed MoE megakernels achieve up to $6\times$ lower latency and $9\times$ higher SM utilization than kernel-based MoE baselines on one multi-GPU server with 8$\times$H100 GPUs~\cite{flashmoe}, and related megakernel systems report comparable gains in single-node configurations~\cite{mpk,comet,triton-dist}. Frontier MoE models, however, cannot be served on a single node. The aggregate expert weights exceed the HBM of any single 8-GPU server, with DeepSeek-V3 activating 37B of 671B parameters per token~\cite{deepseek-v3} and Qwen3-235B and Llama4 Maverick each using 128 experts per layer~\cite{qwen3,llama4}. Production deployments therefore run expert parallelism across multiple nodes, typically 16--64 GPUs spanning 2--16 nodes~\cite{deepseek-v3,pplx-rdma,nvidia-wide-ep}. Existing research leaves open the question of whether the single-node megakernel gains carry over to multi-node deployments.

\subsection{Performance collapse in multi-node weak scaling}
\label{sec:bg:observation}

We answer this question by distributing a state-of-the-art MoE megakernel~\cite{flashmoe} across eight nodes of the NERSC Perlmutter supercomputer~\cite{perlmutter}. Each Perlmutter GPU node contains four NVIDIA A100 GPUs, and each GPU has a dedicated HPE Cassini NIC that exposes RDMA through libfabric~\cite{libfabric}. The NICs connect into a three-level dragonfly fabric, which provides all-to-all connectivity between nodes with at most three hops on the worst-case path. 

To characterize how the megakernel behaves as deployment size grows, we adopt weak scaling, a standard technique in which the per-GPU workload is held constant while GPUs are added, so that the total problem size grows in proportion to the resources provisioned. Under ideal weak scaling, per-iteration latency would stay almost flat because each added GPU contributes compute and network bandwidth proportional to its share of the work, and any deviation from this flat curve isolates overheads that grow with scale rather than with problem size. We apply this method to our Perlmutter deployment by fixing the per-GPU workload (\ie tokens sent per device) and sweeping node count from one to eight. We evaluate scaling megakernels of three state-of-the-art MoE models that span the compute-communication spectrum. Qwen3-30B~\cite{qwen3} is communication-bound, GPT-OSS-120B~\cite{gpt-oss} is balanced, and Llama4-Scout-17B~\cite{llama4} is compute-bound.\footnote{Computed as TFLOPs/GB from per-token FLOPs (including the gated MLP factor $\times 6$) over communication volume. Qwen3 achieves 4.6, GPT-OSS 17.3, and Llama4 49.2.}

\myparab{Slowdown in communication-bound models.} Figure~\ref{fig:motivation-weak-scaling} (top) shows that megakernel performance degrades as we increase node count, and the degradation is worse for models with higher communication-to-compute ratio. Communication-bound Qwen3-30B slows down by up to $10\times$ at 8 nodes relative to the single-node baseline. Balanced GPT-OSS degrades by $1.3$--$1.7\times$ with each doubling of node count. Compute-bound Llama4 scales near-linearly, with only a $1.1$--$1.3\times$ slowdown. This correlation between the slowdown and the communication-to-compute ratio is our first indication that the bottleneck lies on the inter-node communication path rather than in the compute pipeline. SM traces confirm this hypothesis (Figure~\ref{fig:motivation-weak-scaling}, bottom). Processor CTAs stall between tile computations, waiting on signals that arrive late, and the SMs that tile-level parallelism was designed to keep busy end up waiting on the network transport.

The performance collapse is surprising because the megakernel abstraction itself does not change as the deployment grows from one node to many. The scheduler still dispatches on flag flips, and the sender still issues \putsignal into a symmetric buffer. What changes is how the \PUT is realized underneath. Within a node, a \PUT is a write over NVLink that completes at NVLink latency, and prior work has shown NVLink transfers to scale near-linearly with concurrency~\cite{nvlink-eval}. Across nodes, a \PUT traverses an RDMA transport, and the structure of that transport is what we examine next.

\subsection{Device-initiated RDMA and its submission paths}
\label{sec:bg:rdma-proxy}

Remote Direct Memory Access (RDMA) lets a sender transfer data into the memory of a remote machine without involving either machine's CPU on the data path. The sender names the local source and the remote destination memory regions, and the two NICs cooperate to copy the bytes between them. The sending NIC reads the source region from local memory and emits network transfers, and the receiving NIC writes the arriving bytes into the destination region on the remote host. In ML servers these memory regions reside in GPU HBM, and each NIC reaches its local GPU's HBM over PCIe~\cite{gpudirect-rdma}.

\myparab{Submission paths of device-initiated RDMA.} 
Megakernels use \emph{device-initiated} RDMA, in which a GPU thread triggers a transfer from inside a GPU kernel through a call like \texttt{putmem\_signal\_nbi()}. Device-initiated RDMA is a property of the programming model that allows a GPU to be the semantic initiator of the data transfer. A separate axis determines which processor forwards the request to the NIC hardware, which we call the \emph{submission path} of the RDMA transfer. Device-initiated RDMA has two submission paths. In the \emph{GPU-direct} submission path, exemplified by NVIDIA's IBGDA on InfiniBand~\cite{ibgda-ibrc-blog}, GPU threads submit requests to the NIC through a hardware interface the NIC exposes to the GPU, and no CPU thread participates in the process. In the \emph{proxy-based} submission path, GPU threads cannot reach the NIC and instead write requests into a host-resident queue. A dedicated CPU proxy thread drains the queue and forwards each request to the NIC on the GPU's behalf.

\myparab{Most production fabrics rely on proxy-based submission.} The GPU-direct submission path requires NIC hardware support that is absent on several network fabrics, including AWS EFA~\cite{efa}, HPE Slingshot-11~\cite{slingshot}, and Broadcom-based cloud deployments~\cite{uccl-ep}. The NVSHMEM communication library supports two proxy-based transport backends, Libfabric and InfiniBand Reliable Connection (IBRC)~\cite{libfabric,ibgda-ibrc-blog}. A portable multi-node megakernel must work correctly and perform well over both proxy-based and GPU-direct RDMA transport.

\begin{figure}
\centering
\includegraphics[width=\linewidth]{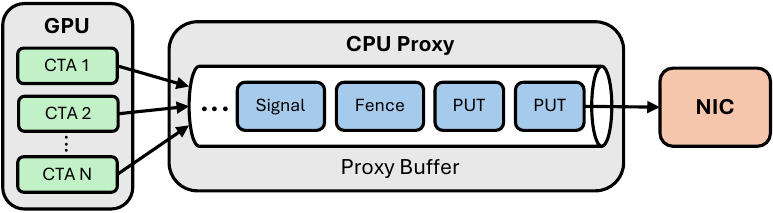}
\caption{\small{Proxy-based RDMA submission path. GPU threads enqueue network requests into a host-resident buffer that a dedicated CPU proxy drains into the NIC. NVSHMEM uses a single proxy channel per PE, and requests from every CTA on a GPU are funneled through one shared submission path.}}
\label{fig:bg-proxy}
\end{figure}

\begin{figure*}[t]
\centering
\includegraphics[width=\textwidth]{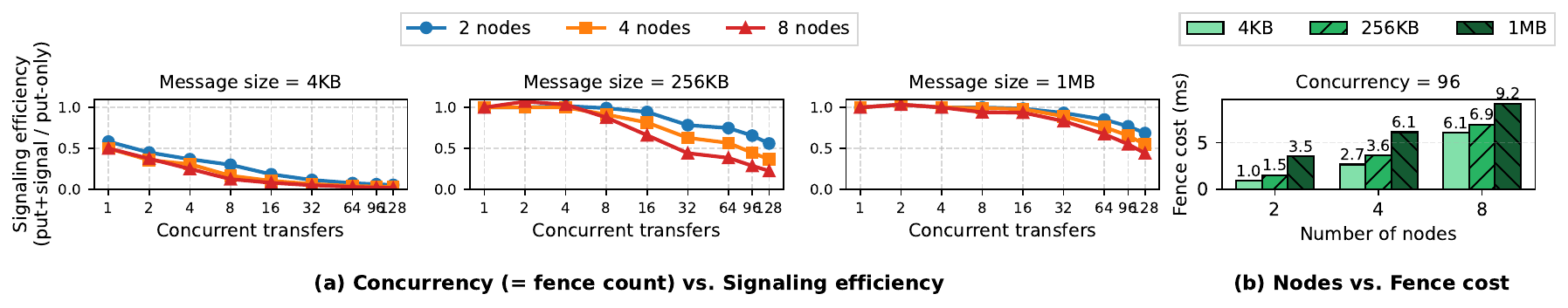}
\includegraphics[width=\textwidth]{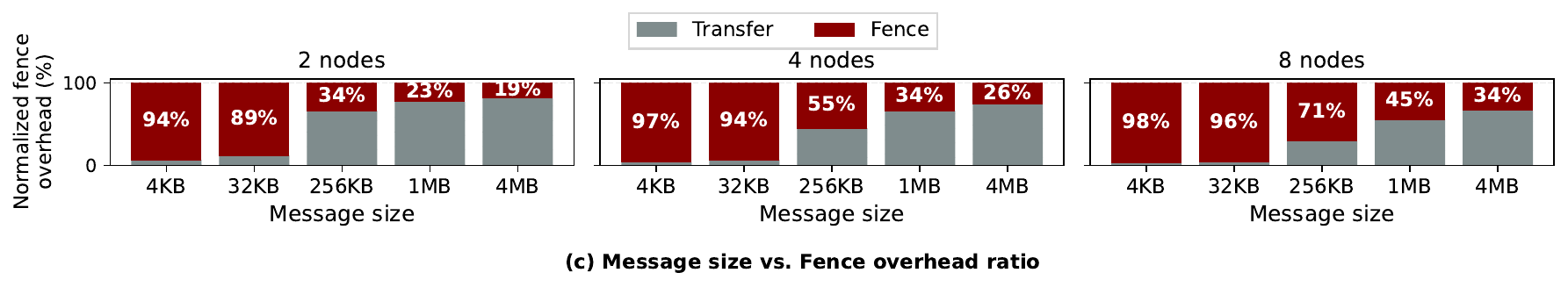}
\caption{\small{Signaling overhead across message sizes, node counts, and concurrency levels. (a) Signaling efficiency, defined as coupled put+signal throughput normalized to pipelined put-only, collapses with increasing concurrency and node count. (b) Fence cost at 96 concurrent transfers grows with both node count and message size. (c) Fence overhead accounts for up to 98\% of total communication time and grows with node count at 96 concurrent transfers. Signal cost is negligible ($<$0.2\% of total time) and included in the transfer portion.}}
\label{fig:motivation-signal-overhead}
\end{figure*}

\myparab{The proxy submission path under megakernel traffic.} Figure~\ref{fig:bg-proxy} shows how a proxy-based transport handles device-initiated requests. The NVSHMEM library, which is the communication library megakernels commonly use, allocates one proxy channel per PE, and every CTA on the GPU writes its requests into the host-resident queue for that channel. The proxy thread drains the queue in FIFO order and forwards one request at a time to the NIC. The number of concurrent requests that traverse this single channel follows from how a megakernel distributes expert computation. With $E$ experts sharded across $P$ PEs, each PE hosts $E/P$ experts. The proxy carries inter-node traffic only, because intra-node transfers use NVLink rather than the NIC, so each PE issues one transfer per \emph{remote} expert per dispatch. Denoting the number of PEs on the sender's own node as $P_{\text{local}}$, a PE issues $(P - P_{\text{local}}) \cdot (E/P)$ concurrent transfers through its proxy channel in the worst case. In our Qwen3-30B configuration at 4 nodes and 16 GPUs, each Perlmutter node hosts 4 PEs, which leaves 12 remote PEs holding 8 experts each, so one PE sends 96 concurrent transfers per dispatch through the single proxy channel. 

\subsection{Root cause: fence-induced serialization}
\label{sec:motivation:root-cause}

Large number of concurrent data transfers required by megakernels (\eg 96 per dispatch to a single proxy channel) would not impose a performance bottleneck if the proxy could pipeline these transfers. We find that it cannot, and the reason lies in how the proxy honors the ordering semantics of \texttt{putmem\_signal\_nbi}. Each \texttt{putmem\_signal\_nbi} call guarantees that when the signal becomes visible at the receiver, the payload of the corresponding \PUT has already been transferred. The proxy enforces this guarantee by expanding the call into a three-step sequence on its channel: a \PUT, a \fence that waits for every prior \PUT on the channel to complete, and the \signal. The fence keeps the \signal behind its \PUT at the NIC to ensure that the data transfer is complete. Within a node the proxy is not on the path and this \fence does not apply, which is why \putsignal carries no observable cost over NVLink. Across nodes, the \fence forces the proxy to stop draining new requests until every in-flight \PUT on the channel has returned a completion from the NIC.

\myparab{Isolating fence cost.} We measure the cost of the fence with a microbenchmark that strips away application-level complexity of the experiment in Figure~\ref{fig:motivation-weak-scaling}. Each PE issues $N$ concurrent RDMA transfers to remote PEs, and we compare a put-only upper bound, which issues \puts without signals, against the coupled \putsignal primitive that megakernels invoke. We sweep concurrency from 1 to 128 transfers and message sizes from 4\,KB to 4\,MB across 2--8 Perlmutter nodes. Figure~\ref{fig:motivation-signal-overhead} summarizes the results. Coupled \putsignal throughput falls to 2\% of the \PUT -only ceiling at 96 concurrent transfers across 8 nodes (Figure~\ref{fig:motivation-signal-overhead}a). The mechanism designed to enable fine-grained overlap becomes the bottleneck that prevents it, and the performance collapse stems from two compounding effects that we isolate next.

\myparab{Concurrent fences interleave in the shared channel.} Concurrent CTAs submit \PUT$\to$\fence$\to$\signal triples into one shared proxy FIFO queue (Figure~\ref{fig:design-cta-nic}a). Each fence waits for every work request ahead of it, including \puts posted by other CTAs, so as concurrency grows each fence encounters more preceding PUTs and the aggregate drain cost climbs (Figure~\ref{fig:motivation-signal-overhead}a). The effect peaks at small message sizes, where fence cost dominates transfer time, and persists at 1\,MB, where efficiency stays below 50\% at 8 nodes (Figure~\ref{fig:motivation-signal-overhead}b).

\myparab{A single fence's drain grows with node count.} A fence waits until every \PUT ahead of it has returned a completion. As the deployment adds nodes, those PUTs target more distinct remote destinations, and the drain waits for the slowest completion across all of them, a tail latency that grows with destination count. Figure~\ref{fig:motivation-signal-overhead}b quantifies this effect. At 96 concurrent transfers, aggregate fence time grows from $0.96$\,ms at 2 nodes to $6.1$\,ms at 8 nodes for 4\,KB messages, and from $3.5$\,ms to $9.2$\,ms for 1\,MB messages. Figure~\ref{fig:motivation-signal-overhead}c shows that fence overhead accounts for up to 98\% of total communication time at small message sizes and remains above 19\% even at 4\,MB. Megakernels operate in this regime, sending many concurrent per-expert transfers so that the receiver can begin dependent compute early at tile granularity (\eg $128\times 64$ BF16 tiles of 16\,KB each). The regime in which megakernels expect to gain the most is therefore the regime in which proxy-based signaling degrades the most.
\subsection{Why strawman solutions fail}
\label{sec:bg:strawman}

Two natural approaches fail to remove the serialization bottleneck while preserving fine-grained overlap. Bulk batching reduces fence count by consolidating many small transfers into fewer large ones, and it does so by reverting to the coarse-grained synchronization that megakernels were designed to eliminate. Raising proxy parallelism through multiple proxy threads or channels removes the single-FIFO bottleneck but requires a cross-channel ordering mechanism to preserve put-with-signal semantics, which reintroduces synchronization with added complexity (\S\ref{sec:discussion}).

These strawmen expose the tension at the heart of the problem. Fine-grained, per-expert signaling is what lets the receiver's scheduler learn tile dependencies at CTA granularity, and coarsening it forfeits the overlap that motivates megakernels. Every fine-grained signal, however, requires a fence, and on a proxy-based RDMA transport every fence serializes the entire submission channel. Removing this bottleneck therefore requires redesigning both the signaling protocol and the proxy transport so that serialization disappears and fine-grained overlap survives.

%% file: tex/design.tex
\section{\sys Design}
\label{sec:design}

\sys eliminates the \fence-induced serialization identified in \S\ref{sec:bg:multinode} through two complementary mechanisms that attack the problem at different layers. At the protocol layer, \emph{decoupled signaling} (\S\ref{sec:design:decouple}) restructures how CTAs submit \puts and \signals so that the proxy FIFO no longer interleaves fences with data transfers.
At the transport layer, \emph{NIC-side ordering} (\S\ref{sec:design:nic}) delegates ordering enforcement to the NIC hardware so the proxy never blocks on a \fence. Together, the two mechanisms remove both the frequency and the per-\fence cost of serialization while preserving the fine-grained \putsignal interface that megakernels depend on.

\begin{figure}[t]
\centering
\includegraphics[width=\linewidth]{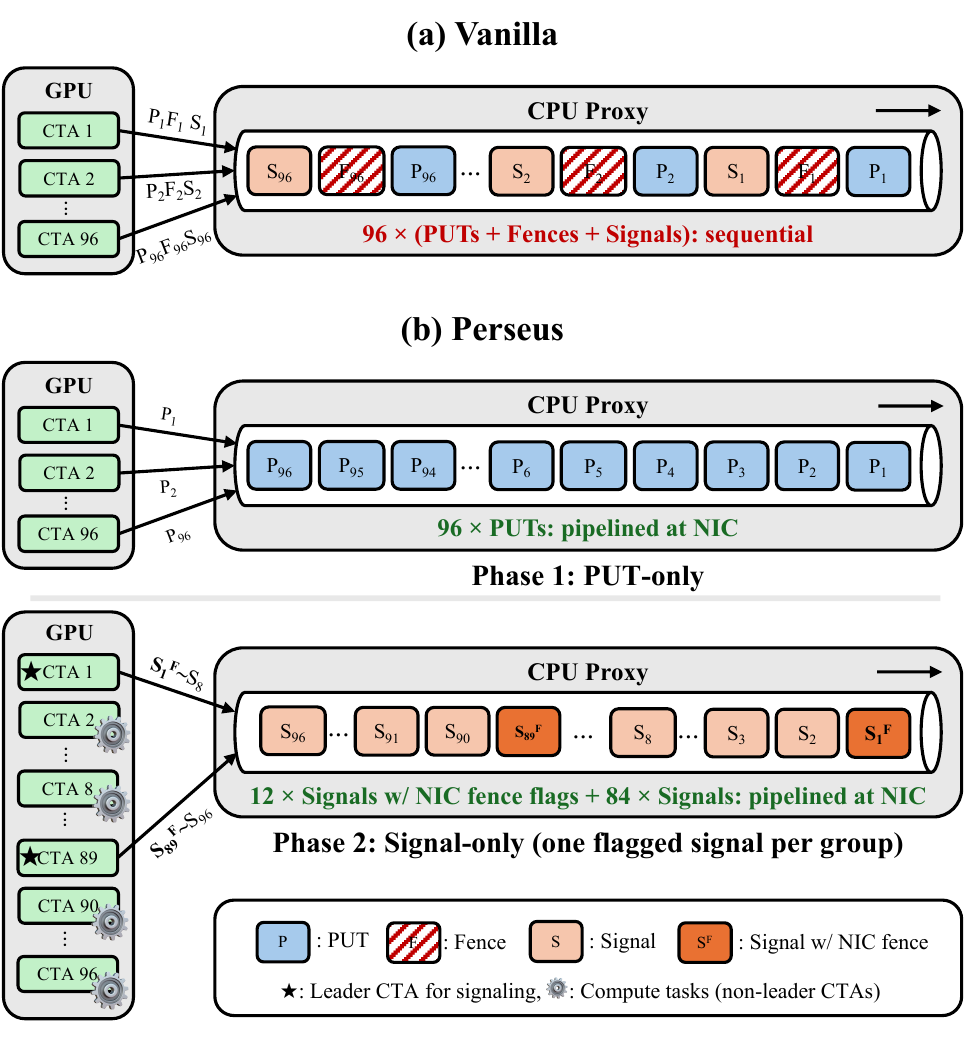}
\caption{CTA-to-proxy interaction for 96 remote experts across 12 remote PEs.
(a) \textbf{Vanilla}: each CTA issues \PUT+\fence+\signal, serializing all 96 transfers.
(b) \textbf{\sys}: \textbf{Phase 1} pipelines 96 \puts back-to-back, which the NIC pipelines freely. \textbf{Phase 2} issues 96 signals, of which only 12 carry NIC fence flags (one per group).
Leader CTAs (\ding{72}) issue flagged \signal; others proceed to compute.}
\label{fig:design-cta-nic}
\end{figure}

\subsection{Decoupled signaling}
\label{sec:design:decouple}
In the coupled \PUT$\to$\fence$\to$\signal sequence that vanilla \putsignal issues, the \fence is the serialization point. Each \fence blocks the proxy until all prior \puts on the channel complete, so the proxy cannot submit new \puts while one \fence is draining. In our running example (Qwen3-30B, 4 nodes, 16 GPUs), each PE dispatches tokens to 96 remote experts and therefore drains the proxy 96 times per dispatch (Figure~\ref{fig:design-cta-nic}a). Merging multiple small transfers into larger ones will amortize the \fence cost but will also shrink transfer concurrency and delay signals, eliminating the tile-level overlap that motivates megakernels (\S\ref{sec:bg:multinode}).

\noindent\textbf{Key insight.}
The \fence is only required before the \signal, not after each \PUT. If all CTAs submit their \puts first and defer fencing until a single leader CTA issues the \signals, the proxy FIFO sees a burst of \puts followed by one \fence followed by a batch of \signals, rather than a tightly interleaved sequence of triples. This reordering delivers two benefits. It allows the NIC to pipeline the \puts concurrently, because no \fence is between them.
It also reduces the number of \fences from one per expert to one per group of experts, because a single \fence now orders all signals in the group.

\begin{algorithm}[t]
\caption{Decoupled signaling}
\label{alg:decoupled}
\begin{algorithmic}[1]
\small
\Require Expert index $e$, group $g = \lfloor e / \textit{group\_size} \rfloor$
\Statex \textbf{Phase 1: Data transfer (all CTAs)}
\State Stage tokens for expert $e$ to local buffer
\State \Call{put\_nbi}{dst, src, size, peer}
  \Comment{no signal}
\State \Call{atomic\_add}{counter[$g$], 1}
  \Comment{notify leader}
\If{$\neg$is\_leader($g$)}
  \State \textbf{resume as schedulable}
    \Comment{free for compute}
\EndIf
\Statex
\Statex \textbf{Phase 2: Signaling (leader CTA only)}
\State \Call{wait\_until}{counter[$g$] $=$ group\_size}
\State \Call{fence}{ }
  \Comment{one per group}
\For{each expert $e'$ in group $g$}
  \State \Call{signal}{sig[$e'$], peer}
\EndFor
\end{algorithmic}
\end{algorithm}

\noindent\textbf{Mechanism.}
\sys realizes this insight with a two-phase protocol (Algorithm~\ref{alg:decoupled}, Figure~\ref{fig:design-cta-nic}b). Phase~1 performs data transfers. Each CTA stages its tokens, issues a non-blocking \PUT without a \signal, and increments a per-group atomic counter to announce completion to the leader (Lines~1--3). Non-leader CTAs then return to the megakernel scheduler and become available for compute tasks, including expert compute for tokens routed locally or over NVLink (Line~5). This overlap of communication with compute improves CTA utilization (\S\ref{sec:eval}). Phase~2 performs signaling. The leader CTA waits until the counter reaches the group size, which guarantees that every \PUT in the group has entered the proxy FIFO (Line~7). It then issues a single \fence followed by the signals for every expert in the group (Lines~8--11).

\noindent\textbf{Correctness.}
Algorithm~\ref{alg:decoupled} preserves the \putsignal ordering by construction. The atomic counter ensures that all \puts in the group enter the proxy FIFO before the leader submits the \fence. The single proxy thread then preserves submission order, so the \fence follows every \PUT in the group. The \fence guarantees that all \puts complete at the NIC before any \signal is processed. By transitivity, every \signal in the group is ordered after every \PUT in the group.

\begin{figure}[t]
\centering
\includegraphics[width=\linewidth]{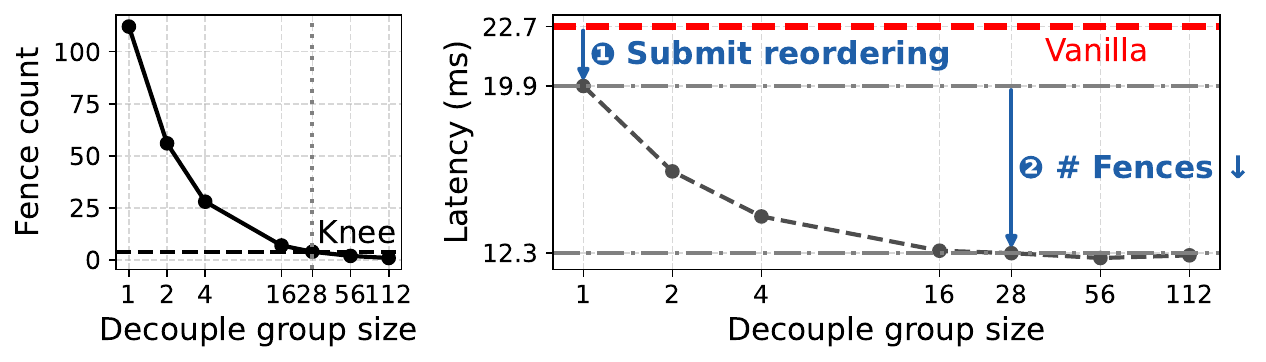}
\caption{Effect of decoupled signaling ($S{=}1K$, 8 nodes).
Left: \fence count \vs group size. Right: latency \vs group size with coupled baseline (dashed red). Contribution to speedup can be decomposed to: \ding{182} submission reordering (coupled$\to$decoupled at group size 1), \ding{183} \fence count reduction from 112 (group size 1)$\to$ 4 (group size 28).}
\label{fig:decouple-granularity}
\end{figure}

\noindent\textbf{Choosing group granularity.}
The size of the group exposes a tradeoff between \fence amortization and intra-\fence waiting. Small groups retain many \fences, while large groups force each \fence to wait on more outstanding \puts. To characterize this tradeoff, we expose group size as a tunable parameter and sweep across all divisors of the remote expert count\footnote{Group size = number of remote experts sharing one fence, \eg group size 4 with 112 remote experts yields 28 fences.}. Figure~\ref{fig:decouple-granularity} shows results at $S{=}1K$ on 8 Perlmutter nodes and separates the total speedup into two components. The first is \emph{submission reordering} (\ding{182}). Even at group size~1, where \fence count matches the coupled baseline, decoupling alone reduces latency by 12\% (22.7$\to$19.9\,ms) because the NIC pipelines \puts back-to-back instead of processing them behind interleaved \fences. The second is \emph{\fence count reduction} (\ding{183}). Increasing group size from 1 to 28 cuts \fences from 112 to 4 and delivers an additional 38\% latency reduction (19.9$\to$12.3\,ms).
Further coarsening has diminishing returns, aligning with the knee in the \fence-count curve (Figure~\ref{fig:decouple-granularity}, left).

\sys defaults to per-PE grouping, in which one group consists of the experts destined for one remote PE. Per-PE grouping is at the knee of the decoupling benefit curve (Figure~\ref{fig:decouple-granularity}, left) and captures most of the \fence reduction benefit while bounding the number of transfers each \fence must wait for. Our sweep confirms that this default is within 2\% of the optimal group size across sequence lengths from 1K to 64K. The exposed configuration allows further tuning for specific workloads and hardware environments.

\subsection{NIC-side ordering}
\label{sec:design:nic}
While decoupled signaling reduces \fence frequency, each remaining \fence still pays the full cost of draining the proxy. A \fence polls a completion counter until all outstanding requests on the channel have returned a completion, and throughout this drain the proxy is idle and cannot submit new requests (Figure~\ref{fig:proxy-nic-timeline}a--b). This drain cost grows with the number of in-flight transfers and with destination count (\S\ref{sec:bg:multinode}).

\noindent\textbf{Key insight.}
Proxy-side drain is not the only way to enforce \putsignal ordering. Modern RDMA NICs expose a per-request \emph{\fence flag} (\eg \texttt{FI\_FENCE} in Libfabric and \texttt{IBV\_SEND\_FENCE} in InfiniBand verbs) that shifts ordering enforcement from software polling into NIC hardware. When the NIC encounters a flagged request, it defers that request until all prior requests on the same connection have completed, tracked through internal hardware registers rather than a software counter. In this way, the NIC absorbs the ordering cost in hardware while the proxy continues submitting requests.

\noindent\textbf{Mechanism.}
\sys replaces the proxy's blocking drain with the NIC \fence flag.
When a \fence is requested, the proxy sets a pending flag rather than polling for completions and continues submitting work. The next \signal sent to the NIC carries the flag, which instructs the NIC to defer that \signal until every prior request on its connection has completed (Figure~\ref{fig:proxy-nic-timeline}c). Thus, the proxy does not stall, and the ordering guarantee is preserved by the NIC itself.

\noindent\textbf{Correctness.}
The NIC \fence flag provides the same ordering guarantee as a proxy-side drain. No operation submitted after the flag can be processed by the NIC before all prior operations on that connection complete. Since the proxy submits \puts before the flagged \signal in submission order, the NIC completes all \puts before processing the \signal. On multi-queue pair transports, peer-hash routing (\S\ref{sec:impl}) pins a peer's \puts and its \signal to the same connection, so per-connection ordering is enough for per-peer correctness.

\begin{figure}[t]
\centering
\includegraphics[width=\linewidth]{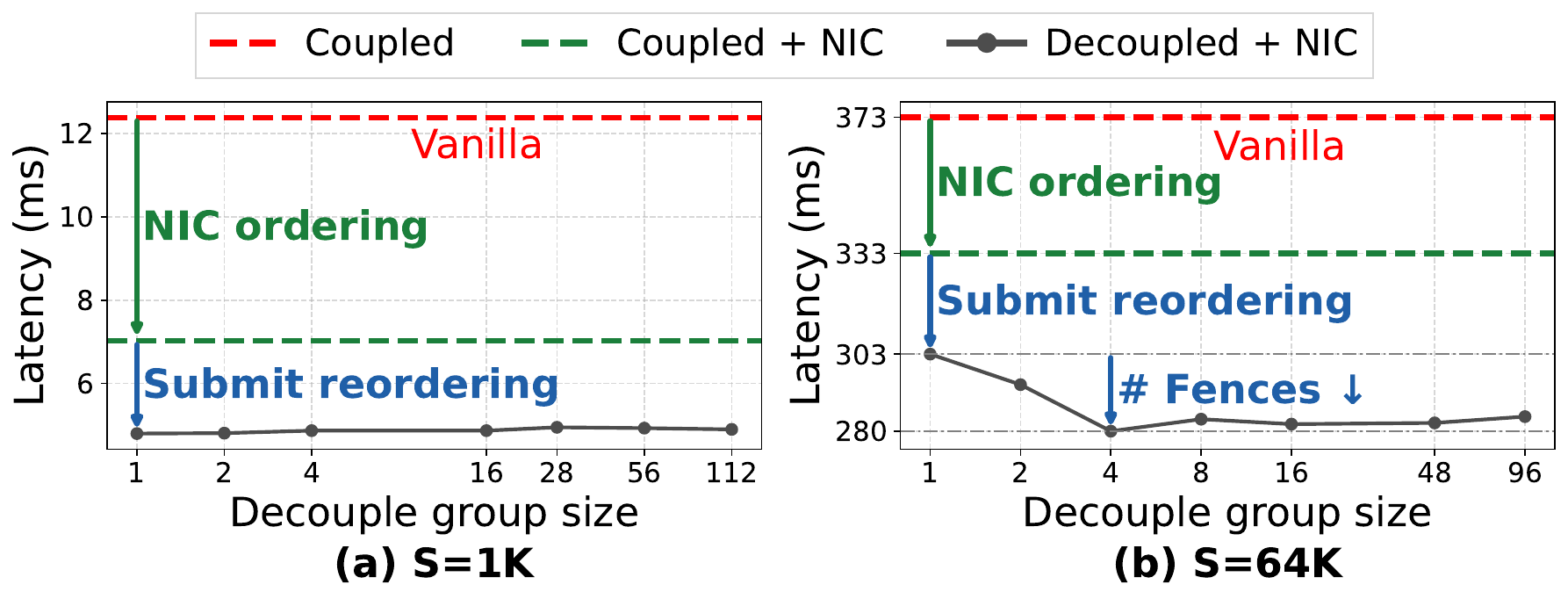}
\caption{Combined effect of decoupled signaling and NIC-side ordering (4 nodes, Qwen3-30B).
Baselines: vanilla (red) and coupled + NIC ordering (green) as horizontal lines; decoupled curves swept across group sizes with NIC-side ordering enabled.
(a)~$S{=}1K$: NIC-side ordering dominates; group size has minimal impact.
(b)~$S{=}64K$: all three effects contribute.}
\label{fig:nic-ordering-effect}
\end{figure}

\subsection{Why the two mechanisms are complementary}
\label{sec:design:combined}

Applied alone to vanilla coupled signaling, NIC-side ordering removes every proxy stop; the NIC absorbs the ordering cost rather than blocking the proxy (Figure~\ref{fig:proxy-nic-timeline}c).
Combined with decoupled signaling, only the first signal per group needs the flag, since the proxy submits signals sequentially and the NIC preserves that order (Figure~\ref{fig:proxy-nic-timeline}d).

Figure~\ref{fig:nic-ordering-effect} exposes the division of labor between the two mechanisms.
At small messages ($S{=}1K$), NIC-side ordering drives per-fence cost to near-zero and flattens sensitivity to group size; the speedup comes from NIC ordering plus submission reordering, and fence count is irrelevant (Figure~\ref{fig:nic-ordering-effect}a).
At large messages ($S{=}64K$), the NIC must defer each flagged \signal until preceding PUTs complete, and larger transfers take longer to finish, so per-fence cost is no longer negligible and fence count reduction regains value (Figure~\ref{fig:nic-ordering-effect}b).
All three effects then contribute to the combined speedup.

The two mechanisms are complementary. Decoupling reduces fence frequency and matters most when fence count dominates. NIC-side ordering eliminates per-fence blocking and matters most when per-fence overhead dominates. Together they cover the full regime from overhead-dominated to transfer-dominated workloads.

%% file: tex/implementation.tex
\section{Implementation}
\label{sec:impl}

\myparab{Decoupled signaling.}
We implement decoupled signaling on top of FlashMoE~\cite{flashmoe}, the state-of-the-art open-source MoE megakernel with multi-node support, modifying only the RDMA request logic and leaving core megakernel components (symmetric tensor layout, subscriber, scheduler, worker modules) untouched. Dispatch to intra-node experts remains unchanged. For inter-node communication, we implement two passes. In the first pass, each CTA stages tokens to the local heap, issues \texttt{nvshmem\_putmem\_nbi()} without signaling, and notifies the per-PE group leader by incrementing an atomic counter (\texttt{cuda::atomic::fetch\_add()}). In the second pass, each group leader CTA calls \texttt{nvshmem\_fence()} followed by \texttt{nvshmemx\_signal\_op()}.

\noindent\textbf{NIC-side ordering.}
Our NIC-side ordering modifies fewer than 100 lines in NVSHMEM's transport modules (v3.5.21). It replaces the proxy's \texttt{quiet}-based drain (\texttt{fi\_cntr\_wait} for Libfabric, \texttt{check\_poll\_avail} for IBRC) with per-request NIC fence flags (\texttt{FI\_FENCE} for Libfabric, \texttt{IBV\_SEND\_FENCE} for IBRC). The change is entirely in the transport shared library (\texttt{nvshmem\_transport\_*.so}). Users swap in the new library without modifying application code or recompiling the host application.\footnote{We plan to upstream this optimization to NVSHMEM.} Any system that uses NVSHMEM as its communication backend inherits the optimization through this swap. For example, we achieve up to 79$\times$ speedup on Triton-distributed's~\cite{triton-dist} \alltoall benchmark without touching its application code (\S\ref{sec:eval:triton-dist}).

\noindent\textbf{Multi-QP transport adaptation.}
On IBRC with multiple Queue Pairs (QPs), the default round-robin policy spreads operations to a given peer across QPs, but \texttt{IBV\_SEND\_FENCE} enforces ordering only within a QP. Dependent \PUT and signal operations therefore may land on different QPs, which breaks the intended ordering. \sys pins all operations targeting a peer to a deterministic QP (\texttt{qp = pe \% num\_qps}). Dependent operations now share a QP and inherit its FIFO ordering, while different peers still spread across QPs to preserve aggregate bandwidth. This design makes \texttt{IBV\_SEND\_FENCE} effective for NIC-side ordering on multi-QP transports.

%% file: tex/eval.tex
\section{Evaluation}
\label{sec:eval}

In this section, we answer the following questions:
\begin{packeditemize}
    \item How much does \sys reduce megakernel latency across transports and scales (\S\ref{sec:eval:e2e})?
    \item What is the contribution of each optimization (\S\ref{sec:eval:ablation})?
    \item Is \sys robust to expert routing skew (\S\ref{sec:eval:skew})?
    \item Is \sys generally applicable (\S\ref{sec:eval:triton-dist})?
\end{packeditemize}

\subsection{Experimental setup}

\noindent\textbf{Hardware environment.}
We evaluate on three transport configurations: (1) Libfabric (proxy-based, Slingshot-11~\cite{slingshot} Cassini NIC) on NERSC Perlmutter~\cite{perlmutter} (4$\times$ A100 per node, up to 16 nodes, 200\,Gb/s, dragonfly), (2) IBRC (proxy-based, ConnectX-7~\cite{cx7} NIC), and (3) IBGDA (GPU-direct, ConnectX-7 NIC) on a commercial GPU cloud (8$\times$ H100 per node, up to 4 nodes, InfiniBand NDR (400\,Gb/s). IBRC uses 4 Queue Pairs per device, a standard multi-QP configuration for bandwidth scaling that requires explicit cross-QP fencing (\S\ref{sec:impl}); IBGDA uses its default configuration.\footnote{We swept Reliable Connection (RC) Queue Pair configuration for IBGDA and checked no variance in latencies.} These configurations cover 64 A100 GPUs (Perlmutter) and 32 H100 GPUs (cloud cluster), spanning typical expert-parallel scale for multi-node MoE inference~\cite{nvidia-wide-ep}.

\noindent\textbf{Base systems.}
Our prototype builds on FlashMoE~\cite{flashmoe} and NVSHMEM~\cite{nvshmem} (v3.5.21). Unmodified versions are referred to as \emph{vanilla}. \sys modifies FlashMoE's signaling protocol and NVSHMEM's transport layer. We also evaluate on Triton-distributed for generality (\S\ref{sec:eval:triton-dist}).

\begin{table}[h]
\small
\centering
\caption{MoE model configurations. H: hidden dimension, I: intermediate dimension, E: experts, k: top-k.}
\begin{tabular}{lcccc}
\toprule
Model & H & I & E & k \\
\midrule
Qwen3-30B-A3B~\cite{qwen3} & 2048 & 768 & 128 & 8 \\
DeepSeek-V3~\cite{deepseek-v3} & 7168 & 2048 & 256 & 8 \\
GPT-OSS-120B~\cite{gpt-oss} & 2880 & 2880 & 128 & 4 \\
\bottomrule
\end{tabular}
\label{tab:model-config}
\end{table}

\begin{figure*}[t]
\centering
\includegraphics[width=\textwidth]{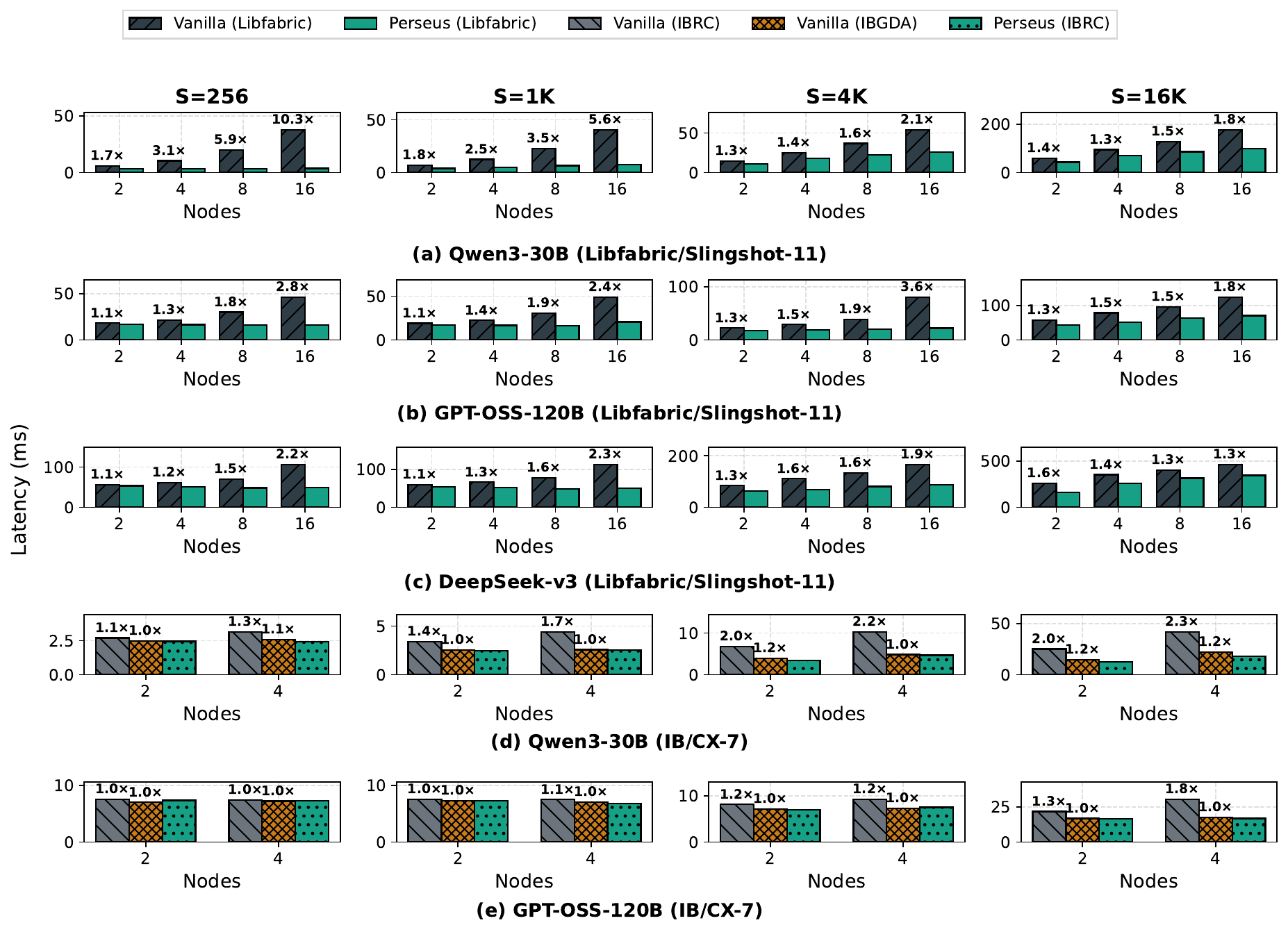}
\caption{End-to-end forward latency: vanilla FlashMoE vs. \sys.
(a)--(c)~Perlmutter (A100, Slingshot-11, Libfabric), 2--16 nodes (8--64 GPUs): \sys achieves up to 10.3$\times$ speedup.
(d)--(e)~Commercial GPU cloud (H100, ConnectX-7, InfiniBand), 2--4 nodes (16--32 GPUs): \sys on IBRC achieves up to 2.47$\times$\protect\footnotemark{} while matching or exceeding IBGDA GPU-direct. Annotations show speedup of \sys over vanilla.}
\label{fig:eval-e2e-latency}
\end{figure*}

\noindent\textbf{Workloads.}
We evaluate three MoE models with varying compute-to-communication ratios (Table~\ref{tab:model-config}). Sequence lengths ($S$) range from 256 to 64K tokens\footnote{We report $S \in \{256, 1K, 4K, 16K\}$ in Figure~\ref{fig:eval-e2e-latency} due to space constraints; results up to $S{=}64K$ are discussed in text.}, covering both overhead-dominated (small $S$) and transfer-dominated (large $S$) regimes. We use balanced expert routing with expert capacity $EC = S \times k / E$ and evaluate skewed routing in \S\ref{sec:eval:skew}.

\subsection{End-to-end performance}
\label{sec:eval:e2e}
\noindent\textbf{Overview.}
Figure~\ref{fig:eval-e2e-latency} shows that \sys achieves up to 10.3$\times$ speedup on Libfabric and up to 2.47$\times$ speedup on IBRC. On IBRC, \sys matches or exceeds vanilla IBGDA GPU-direct (up to 1.2$\times$). The relative performance of proxy-based and GPU-direct transports depends on workload: at per-expert message sizes in our workloads (45\,KB--28\,MB), proxy submission rate is not the bottleneck~\cite{ibgda-ibrc-blog}. Meanwhile, IBGDA consumes SM cycles for NIC submission, which in megakernels competes with concurrent compute; the CPU proxy avoids this interference but is bottlenecked by fence-induced serialization. \sys removes this bottleneck, allowing proxy-based IBRC to match or exceed GPU-direct performance.

\footnotetext{IBRC speedup continues to grow at larger $S$, reaching 2.47$\times$ at $S{=}64K$ on 4 nodes (Qwen3).}

\noindent\textbf{Correctness.}
All results on \sys pass correctness verification with error rate below 1\% across all configurations, matching vanilla error rates under BF16 precision.

\begin{figure*}[t]
\centering
\begin{minipage}[c]{0.7\textwidth}
\includegraphics[width=\textwidth]{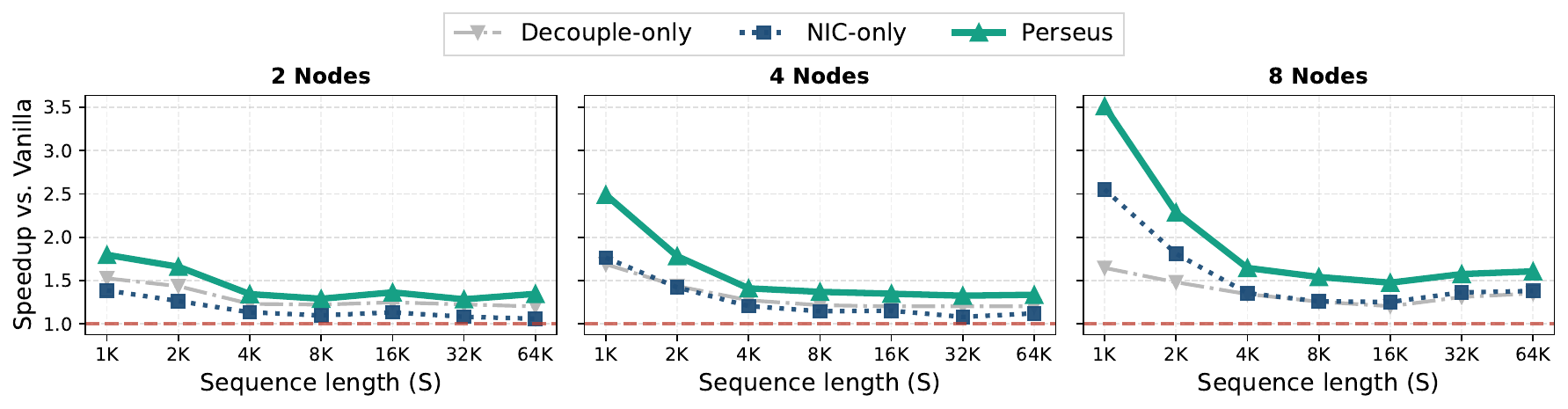}
\end{minipage}
\hfill
\begin{minipage}[c]{0.29\textwidth}
\centering
\scriptsize
\begin{tabular}{l@{\hskip 4pt}c@{\hskip 4pt}c@{\hskip 4pt}c}
\toprule
 & \# Fence & Cost / Fence & Spdup \\
\midrule
Vanilla & 112 & \textcolor{red}{$\uparrow$} & 1.0$\times$ \\
NIC     & 112 & \textcolor{green!50!black}{$\downarrow$} & 1.3--2.6$\times$ \\
Decoup. & 28  & \textcolor{red}{$\uparrow$} & 1.2--1.6$\times$ \\
\textbf{\sys}    & 28  & \textcolor{green!50!black}{$\downarrow$} & \textbf{1.5--3.5$\times$} \\
\bottomrule
\end{tabular}
\vspace{4pt}
\\ \small 8-node ablation breakdown
\end{minipage}
\caption{Left: Ablation of speedup over vanilla for Qwen3-30B on Perlmutter.
At larger node counts, per-fence cost increases and fence reduction is proportionally smaller, so NIC-side ordering contributes more; at smaller node counts, the opposite holds.
Right: 8-node breakdown showing fence count and per-fence cost (\textcolor{red}{$\uparrow$} = full pipeline drain, \textcolor{green!50!black}{$\downarrow$} = lightweight NIC-side flag).}
\label{fig:eval-speedup-breakdown}
\end{figure*}

\noindent\textbf{Speedup across nodes.}
Recall the poor weak-scaling performance of vanilla on Qwen3 in Figure~\ref{fig:motivation-weak-scaling}. \sys enables near-linear scaling across models and transports, achieving higher speedups at larger node counts. As identified in \S\ref{sec:motivation:root-cause}, vanilla's overhead grows with node count because more remote peers increase fence frequency and proxy stalls.

\noindent\textbf{Speedup across workloads.}
Speedup is generally higher on communication-bound configurations: on Libfabric, peak speedup is 10.3$\times$ for Qwen3 vs.\ 2.8$\times$ for GPT-OSS and 2.2$\times$ for DeepSeek-V3. At larger $S$, speedups across models converge as per-transfer cost dominates. Speedup trends also differ across transports: on Libfabric, speedups peak at small $S$; on IBRC, speedups grow with $S$, reflecting different fence cost structures (Appendix~\ref{appdx:alpha-beta}).

\begin{figure*}[t]
\centering
\includegraphics[width=\textwidth]{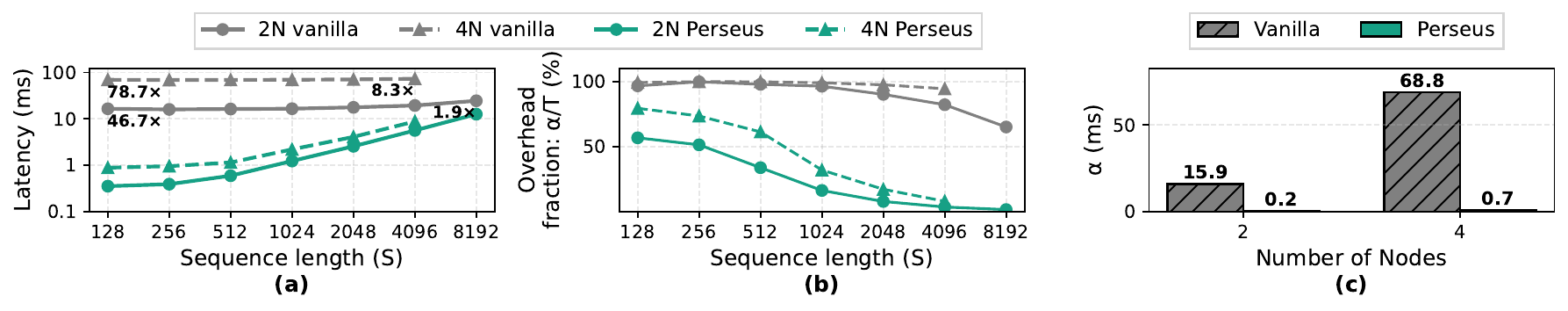}
\caption{Triton-distributed \alltoall microbenchmark (H=2048, Perlmutter).
(a) Latency: vanilla is flat regardless of message size.
(b) Overhead fraction ($\alpha/T$): vanilla remains above 65\%.
(c) $\alpha$ comparison: \sys achieves ${\sim}$99\% reduction in serialization overhead.}
\label{fig:eval-triton-distributed}
\end{figure*}

\begin{figure}[t]
\centering
\includegraphics[width=\linewidth]{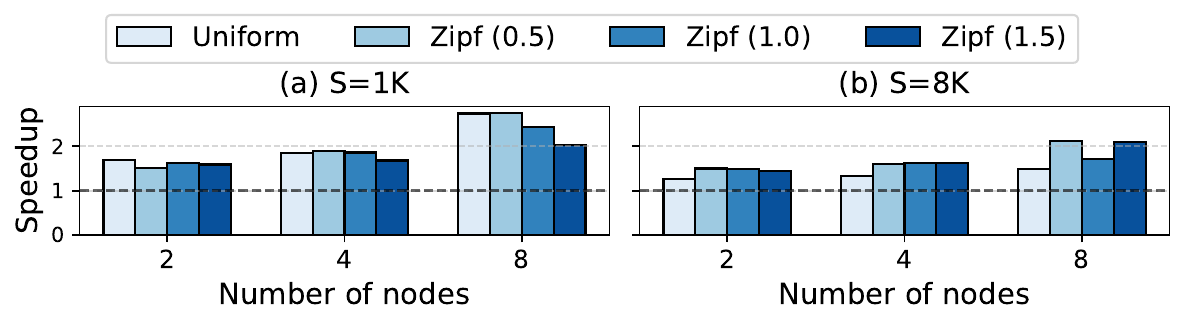}
\caption{Robustness to skewed expert routing (Qwen3-30B, 2--8 nodes on Perlmutter).
Speedup of \sys over vanilla under Zipf-distributed~\cite{zipf} token-to-expert routing with increasing skew exponent (0--1.5).}
\label{fig:eval-skew-robustness}
\end{figure}

\subsection{Ablation}
\label{sec:eval:ablation}

We compare three configurations: (1) decoupled signaling only, (2) NIC-side ordering only, and (3) the full \sys. Figure~\ref{fig:eval-speedup-breakdown} shows results across node counts. At 2 nodes, decoupled signaling (1.2--1.5$\times$) outperforms NIC-side ordering (1.1--1.4$\times$): per-fence cost is small at 2 nodes, so reducing fence count (64 to 4) has more impact. At 8 nodes, the relationship reverses: NIC-side ordering (1.3--2.6$\times$) surpasses decoupled signaling (1.2--1.6$\times$), as the 28 remaining fences are each expensive. Combined, \sys achieves 1.5--3.5$\times$ speedup at 8 nodes, confirming that the two techniques are complementary: decoupled signaling reduces fence \emph{frequency}, while NIC-side ordering reduces per-fence \emph{cost}.

\subsection{Robustness to expert routing skew}
\label{sec:eval:skew}
We inject Zipf-distributed~\cite{zipf} routing with skew exponent from 0 (uniform) to 1.5 (top 10 out of 128 experts receive 82\% of tokens), with expert capacity set to avoid token drops. Figure~\ref{fig:eval-skew-robustness} shows \sys maintains substantial speedup across all skew levels on Qwen3-30B (2--8 nodes): at $S{=}1K$, speedup ranges from 2.7$\times$ to 2.0$\times$ at 8 nodes; at $S{=}8K$, speedup increases with skew (1.5$\times$ at uniform to 2.1$\times$ at Zipf~1.5), consistent with \sys's per-byte cost advantage amplifying under larger transfers (Appendix~\ref{appdx:alpha-beta}).

\begin{figure}[t]
\centering
\includegraphics[width=\linewidth]{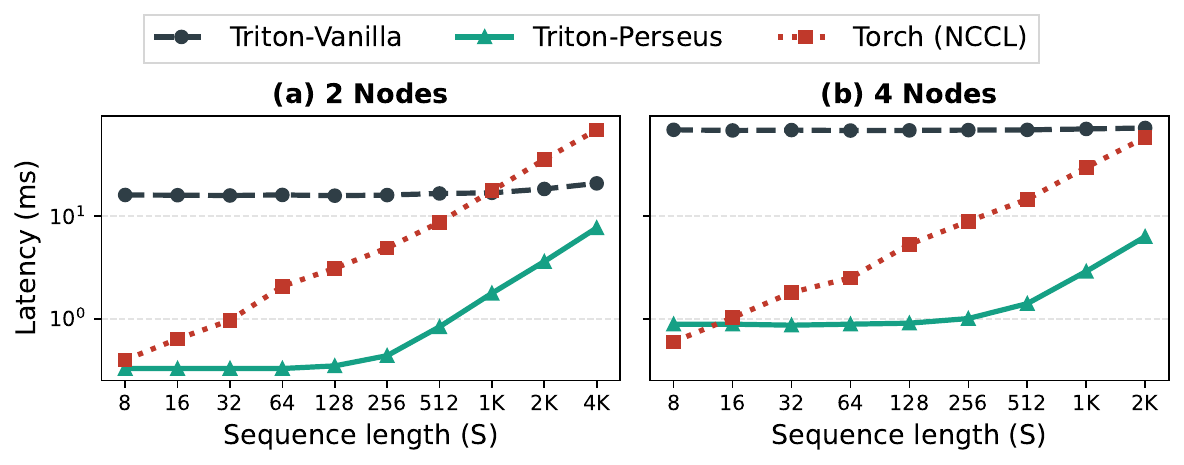}
\caption{GPU-initiated \alltoall of Triton-distributed vs.\ NCCL collective \alltoall ($H{=}2048$, Perlmutter). Triton-vanilla suffers from serialization overhead, losing to NCCL at $S\leq512$. \sys eliminates this overhead, making GPU-initiated \alltoall up to 11$\times$ faster than NCCL. 4-node results limited to $S{=}2048$ due to memory constraints.}
\label{fig:eval-triton-torch}
\end{figure}

\subsection{Generality: Triton-distributed case study}
\label{sec:eval:triton-dist}

We apply NIC-side ordering to Triton-distributed~\cite{triton-dist}'s \alltoall kernel, a communication-only workload with no compute to overlap.\footnote{\texttt{Mega-EP}, Triton-distributed's megakernel, only supports intra-node.} Figure~\ref{fig:eval-triton-distributed} shows that vanilla latency stays nearly flat regardless of message size, with $\alpha$ accounting for 65--100\% of total latency. \sys reduces $\alpha$ by ${\sim}$99\% (68.8 $\to$ 0.7\,ms at 4 nodes), bringing latency close to transfer-dominated behavior. Without modifying application code, this yields up to 79$\times$ speedup (59.6$\times$ on average at 4 nodes).

\noindent\textbf{Comparison with NCCL.}
Without \sys, GPU-initiated \alltoall is on average 18.7$\times$ slower than NCCL (Figure~\ref{fig:eval-triton-torch}).\footnote{4-node results limited to $S{=}2$K due to memory constraints.} \sys makes it up to 11$\times$ faster than NCCL, restoring the latency advantage of fine-grained communication at small message sizes. The scaling behaviors contrast: NCCL's collective overhead grows with data volume, while \sys incurs near-zero fixed overhead, demonstrating that once serialization is removed, GPU-initiated communication consistently outperforms bulk-synchronous collectives in multi-node deployment.

\begin{figure}[t]
\centering
\includegraphics[width=\linewidth]{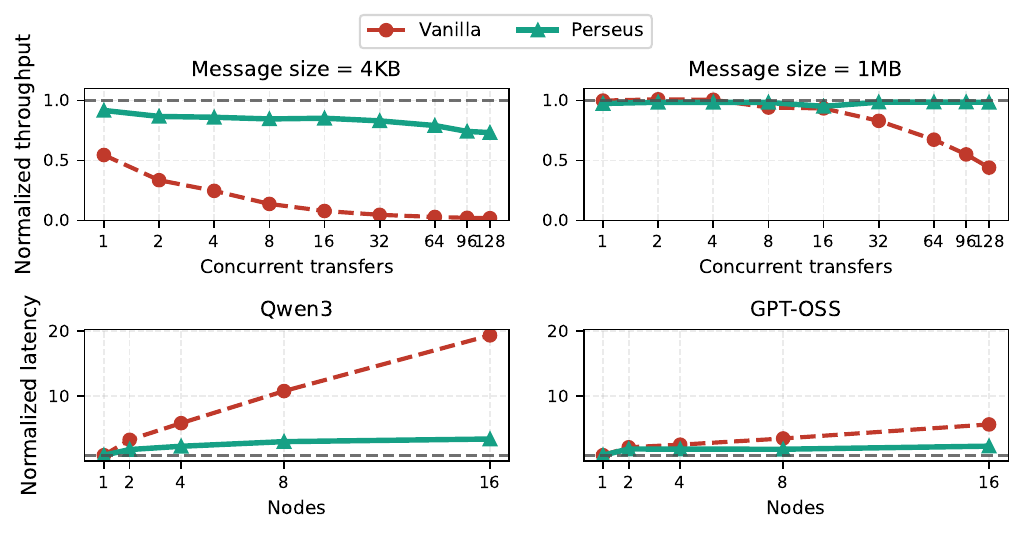}
\caption{Scalability recovery with \sys. 
Top: microbenchmark throughput at 8 nodes, revisiting Figure~\ref{fig:motivation-signal-overhead}a. 
Bottom: end-to-end weak scaling normalized to 1-node baseline ($S$=1024), revisiting Figure~\ref{fig:motivation-weak-scaling}.}
\label{fig:eval-recovery}
\end{figure}

\subsection{Scalability recovery}
\label{sec:eval:recovery}

We revisit the microbenchmark from Figure~\ref{fig:motivation-signal-overhead}a with \sys applied (Figure~\ref{fig:eval-recovery} top). At 96 concurrent transfers with 4\,KB messages, \sys recovers throughput from 2\% to 74\% of the \PUT-only baseline; at larger messages, \sys matches the baseline. Figure~\ref{fig:eval-recovery} (bottom) shows \sys reduces Qwen3's 16-node weak scaling degradation from 19$\times$ to 3.5$\times$ and achieves near-flat scaling on GPT-OSS. \sys also restores TensorCore utilization to 95--98\% of single-node levels (Table~\ref{tab:tc-util}), up from 31\% under vanilla for Qwen3.

\begin{table}[t]
\centering
\small
\caption{TensorCore utilization at 4 nodes ($S{=}1K$), normalized to each workload's single-node execution (100\% = single-node utilization).}
\label{tab:tc-util}
\begin{tabular}{lcc}
\toprule
 & Vanilla & \sys \\
\midrule
GPT-OSS-120B & 75\% & 98\% (+23\%p) \\
Qwen3-30B    & 31\% & 95\% (+64\%p) \\
\bottomrule
\end{tabular}
\end{table}

%% file: tex/related.tex
\section{Related Work}
\label{sec:related}

\noindent\textbf{Systems for distributed MoE.}
Collective-based systems~\cite{megatron,deepspeed-moe,megascale-infer,tutel,fastermoe,janus,lina,megascale-moe,schemoe} optimize EP based on \alltoall primitives. GPU-initiated EP libraries~\cite{deepep,rocm-deepep,mooncake-ep,hybridep,nccl-ep} replace collectives with GPU-initiated point-to-point communication for fine-grained overlap. Recent work optimizes kernels to further increase the overlap~\cite{comet,flux,lancet,tilelink,t3,amd-fused,chimera,flashoverlap}. \sys is complementary to these systems: it addresses transport-level serialization that arises when GPU-initiated communication traverses proxy-based RDMA, a bottleneck that existing systems inherit from the underlying transport.

\noindent\textbf{Megakernels.}
FlashMoE~\cite{flashmoe}, Spector et al.~\cite{hazy-megakernel}, MPK~\cite{mpk}, and Triton-distributed~\cite{triton-dist} fuse LLM operators into persistent GPU kernels for low-latency inference, achieving strong single-node performance but not addressing multi-node scaling. \sys complements this work by characterizing and optimizing transport behavior at multi-node scale.

\noindent\textbf{GPU-initiated communication.}
NVSHMEM~\cite{nvshmem}, NCCL device APIs~\cite{nccl-gin}, NCCLX~\cite{ncclx}, and MSCCL++~\cite{msccl++} provide building blocks for GPU-initiated communication. TransferEngine~\cite{pplx-rdma} and UCCL-EP~\cite{uccl-ep} target portable inter-node communication across heterogeneous GPUs and NICs, avoiding sender-side fence overhead through WriteImm-based notification~\cite{rdma-writeimm}, but require a different interface from the memory-based \putsignal model megakernels rely on (\S\ref{sec:discussion}). \sys instead preserves this interface and eliminates fence-induced serialization within it, enabling drop-in adoption without application or interface changes.

%% file: tex/discussion.tex
\section{Discussion}
\label{sec:discussion}

\noindent\textbf{Why memory-based put+signal?}
Receiver CTAs detect data arrival by polling GPU memory with a single load instruction; alternatives such as CQ polling~\cite{pplx-rdma,uccl-ep} add latency that undermines fine-grained overlap at megakernel level.

\noindent\textbf{Multi-threaded proxy.}
Multiple proxy threads enable concurrent submission but do not remove the ordering requirement. A single thread enforces this via FIFO drain; multiple threads require cross-thread synchronization that reintroduces serialization. Even with one thread, multi-QP IBRC requires cross-QP drains (\S\ref{sec:eval:e2e}). \sys addresses ordering cost directly, regardless of thread or connection count.

\noindent\textbf{Broader applicability.}
\sys applies to any GPU-initiated \putsignal over proxy-based transports; the bottleneck is architectural, not implementation-specific. We demonstrate this across two transports (Libfabric on Slingshot, IBRC on InfiniBand), using only standard RDMA APIs. Decoupled signaling applies to any megakernel using per-transfer \putsignal. Our Triton-distributed case study achieves 79$\times$ speedup without application changes (\S\ref{sec:eval:triton-dist}). Integration with serving engines (\eg vLLM~\cite{vllm}, SGLang~\cite{sglang}) and training frameworks (\eg Megatron~\cite{megatron}) is future work.

\noindent\textbf{Prefill vs. decode.}
Our sweep on $S$ spans both regimes: small $S$ resembles decode (overhead-dominated), while large $S$ resembles prefill (transfer-dominated). \sys improves both: up to 10.3$\times$ at small $S$ where $\alpha$ dominates, and 1.8$\times$ at large $S$ where $\beta$ reduction remains beneficial (Figure~\ref{fig:eval-e2e-latency}).

\noindent\textbf{GPU-direct transports.}
\sys's decoupled signaling with adaptation yields modest gains on IBGDA (up to 1.25$\times$), where proxy serialization is absent (Appendix~\ref{appdx:ibgda}).

%% file: tex/conclusion.tex
\section{Conclusion}
\label{sec:conclusion}

We present \sys, a system that eliminates hidden serialization in multi-node megakernel communication. By decoupling signaling from data transfers and delegating ordering to NIC hardware, \sys achieves up to 10.3$\times$ speedup on Libfabric and 2.47$\times$ on IBRC, matching GPU-direct performance without requiring GPU-direct hardware support. Our results demonstrate that the proxy-based transport architecture is not inherently inferior for fine-grained GPU-initiated communication in megakernels; the bottleneck was in the ordering mechanism, not the proxy itself. As megakernel designs scale to larger deployments and new RDMA fabrics, \sys's transport-level optimizations provide a foundation for efficient multi-node GPU-initiated communication.

%% file: tex/appendix.tex
\section*{Appendix}

\section{\texorpdfstring{$\alpha$-$\beta$}{α-β} analysis: latency decomposition}
\label{appdx:alpha-beta}

\begin{figure}[h!]
\centering
\includegraphics[width=\linewidth]{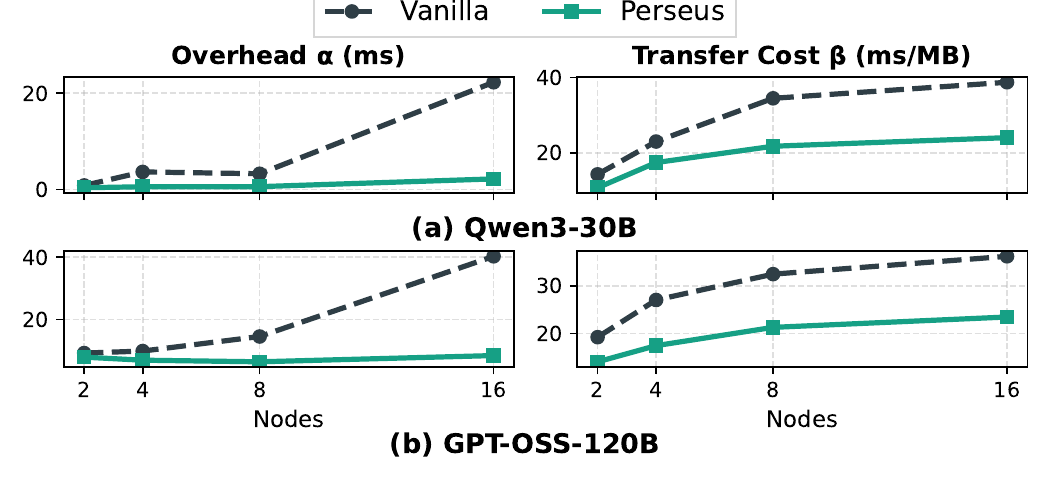}
\caption{Latency decomposition into overhead ($\alpha$) and transfer cost ($\beta$) on Libfabric.
Left: vanilla $\alpha$ grows with node count; \sys stays flat. 
Right: \sys also reduces $\beta$ by 25--38\%.}
\label{fig:eval-alpha-beta}
\end{figure}

\noindent\textbf{Computing $\alpha$-$\beta$ cost.}
To better understand where our performance gains come from, we decompose communication latency using the $\alpha$-$\beta$ model~\cite{alpha-beta}.
The cost to send a message of size $M$ is: $T = \alpha + \beta \times M$, where $\alpha$ captures fixed overhead independent of message size and $\beta$ is the per-byte transfer cost.
We compute $\alpha$ and $\beta$ of both vanilla and \sys across node configurations by fitting linear regression using our sweep results on multiple $S$ values ($R^2 > 0.99$ for all configurations).
We compute $M = EC \times H \times 2$ bytes where $EC = S \times k / E$.
In our setup, $M = S \times 256$ bytes for Qwen3 and $M = S \times 180$ bytes for GPT-OSS.

\noindent\textbf{Latency decomposition.}
Our $\alpha$-$\beta$ decomposition demonstrates \sys reduces both fixed overhead ($\alpha$) and per-byte transfer cost ($\beta$), with different transports exhibiting different reduction profiles.
On Libfabric, $\alpha$ reduction dominates: vanilla's $\alpha$ grows with node count due to increasing fence overhead, while \sys's $\alpha$ remains nearly flat (Figure~\ref{fig:eval-alpha-beta}).
For Qwen3, \sys reduces $\alpha$ by up to 90\% (22.28\,ms $\to$ 2.21\,ms at 16 nodes).
For GPT-OSS, the reduction is up to 79\% (40.3\,ms $\to$ 8.37\,ms at 16 nodes).
On IBRC, $\alpha$ is inherently small (1--5\,ms) because hardware completion queue polling is lightweight compared to Libfabric's software counter drain.
Instead, the multi-QP drain cost inflates $\beta$: \sys reduces $\beta$ by up to 60\% on Qwen3 and 44\% on GPT-OSS, closely matching or even outperforming IBGDA's $\beta$ across all configurations.
Eliminating proxy drains allows the NIC to pipeline transfers without waiting for prior completions, reducing per-transfer cost that scales with message size.

\noindent\textbf{Explaining speedup trends across sequence lengths.}
The $\alpha$-$\beta$ decomposition explains the different speedup trends observed across transports.
The speedup of \sys over vanilla can be expressed as:
$\frac{\alpha_v + \beta_v \cdot M}
      {\alpha_b + \beta_b \cdot M}$
where subscripts $v$ and $b$ denote vanilla and \sys respectively.
On Libfabric, where $\alpha_v \gg \alpha_b$ and the $\beta$ gap is moderate, speedup is highest at small $S$ (small $M$) where $\alpha$ dominates, and decreases at larger $S$ as $\beta \cdot M$ dominates both numerator and denominator.
On IBRC, where $\alpha_v \approx \alpha_b$ but $\beta_v \gg \beta_b$, speedup grows with $S$: at small $M$ the similar $\alpha$ values yield modest speedup, while at large $M$ the $\beta$ gap widens the advantage, peaking at $S{=}64K$ (2.47$\times$ for Qwen3, 1.70$\times$ for GPT-OSS).

\section{Optimizing GPU-direct transports}
\label{appdx:ibgda}
While we focus on proxy-based transports, we briefly examine IBGDA, which avoids proxy serialization via NIC-level in-QP ordering.
With RC QPs (one per peer), \sys's decoupled signaling applies directly: \PUT and \signal WQEs are issued to the same QP, and in-QP ordering ensures correctness without \fences.
Decoupling changes the submission pattern: pipelining \puts before \signals, allowing the NIC to process transfers without interleaved ordering points.
Additionally, we enable warp-parallel signaling within each group, where multiple warps issue signals concurrently.
This optimization is specific to GPU-direct transports, as proxy-based transports serialize all signals through a single proxy thread.
In preliminary evaluation (2 nodes, 16 H100 GPUs), this yields up to 1.25$\times$ speedup on Qwen3 (1.06$\times$ mean).
While modest, these gains are notable on a transport that already avoids proxy serialization.
Further analysis and large-scale evaluation are future work.